\documentclass[aps, pra, a4paper, 10pt, showpacs, twocolumn]{revtex4-1}
\usepackage{amsmath, amssymb, amsthm, bm, textcomp}
\usepackage[T1]{fontenc}
\usepackage[latin9]{inputenc}
\usepackage{graphicx,graphics}
\usepackage{geometry}
\usepackage{color}
\newcommand{\ket}[1]{\left|{#1}\right\rangle}
\newcommand{\bra}[1]{\left\langle{#1}\right|}

\newcommand{\be}{\begin{equation}}
\newcommand{\ee}{\end{equation}}
\newcommand{\eea}{\end{eqnarray}}
\newcommand{\bea}{\begin{eqnarray}}

\begin{document}

\title{Quantum repeaters based on trapped ions with decoherence free subspace encoding}

\author{M. Zwerger$^{1}$, B. P. Lanyon$^{2,3}$, T. E. Northup$^{2}$, C. A. Muschik$^{3,4}$,  W. D\"ur$^{4}$, and N. Sangouard$^{1}$}

\affiliation{
$^1$ Departement Physik, Universit\"at Basel, Klingelbergstra{\ss}e 82, 4056 Basel, Switzerland\\
$^2$ Institut f\"ur Experimentalphysik, Universit\"at Innsbruck, Technikerstra{\ss}e 25, A-6020 Innsbruck,  Austria \\
$^3$ Institut f\"ur Quantenoptik und Quanteninformation der \"Osterreichischen Akademie der Wissenschaften, A-6020 Innsbruck, Austria\\
$^4$ Institut f\"ur Theoretische Physik, Universit\"at Innsbruck, Technikerstra{\ss}e 21a, A-6020 Innsbruck,  Austria}
\date{\today}

\begin{abstract}
Quantum repeaters provide an efficient solution to distribute Bell pairs over arbitrarily long distances. While scalable architectures are demanding regarding the number of qubits that need to be controlled, here we present a quantum repeater scheme aiming to extend the range of present day quantum communications that could be implemented in the near future with trapped ions in cavities. We focus on an architecture where ion-photon entangled states are created locally and subsequently processed with linear optics to create elementary links of ion-ion entangled states. These links are then used to distribute entangled pairs over long distances using successive entanglement swapping operations performed using deterministic ion-ion gates. We show how this architecture can be implemented while encoding the qubits in a decoherence free subspace to protect them against collective dephasing. This results in a protocol that can be used to violate a Bell inequality over distances of about 800 km assuming state of the art parameters. We discuss how this could be improved to several thousand kilometers in future setups. 
\end{abstract}
%



\maketitle

\section{Introduction}
The ability to establish entanglement over long distances is crucial for applications like quantum key distribution \cite{BB84,E91,Scarani09}, distributed quantum computation \cite{Meter06,Meter08,Beals20120686}, reference frame alignment \cite{BartlettRMP,Islam2014}, longer-baseline telescopes \cite{Gottesman2012}, blind quantum computation \cite{Broadbent2009}, and more generally, quantum networks \cite{Kimble08}. In establishing long-distance entanglement, the main difficulties to overcome are loss of qubits and decoherence. Due to the no-cloning theorem \cite{Wootters82} one cannot use redundant encoding or amplification as in the classical case. Quantum repeater protocols for scalable communication circumvent these problems by making use of either quantum error correction \cite{Knill96} or entanglement purification and swapping \cite{Briegel98} (or combinations thereof). In the first case \cite{Knill96} one encodes quantum information in a quantum error correcting code, sends it through the channel and applies quantum error correction at properly spaced quantum repeater stations. In the second case \cite{Briegel98}, one first creates entanglement between neighboring repeater stations and then iteratively applies entanglement swapping and purification. Both schemes require the control of a number of qubits at each node which is above what can be handled with present day technology \cite{Briegel98,Duer1999,Mu16}. 

This requirement motivated people to study simpler quantum repeaters without entanglement purification which are not scalable \footnote{a quantum repeater is scalable if one can create a Bell pair with a fidelity exceeding a predefined value over arbitrarily long distances with at most polynomially growing resources}, but might beat the direct transmission of photons through fibers. These architectures use a reduced number of qubits and a few entanglement swapping operations so that the errors are small enough to allow quantum communication without entanglement purification. Notice that the term quantum repeater originally referred to scalable protocols but is now commonly used for any long-range quantum communication scheme.  Many proposals are based on atomic ensembles and linear optics following the initial work of \cite{Duan2001}; for a review see \cite{Sangouard11}. In \cite{Sangouard09} a scheme based on single trapped ions was put forward which achieved higher rates than with atomic ensembles, mostly due to deterministic entanglement swapping operations in this setup (see also \cite{DuanRMP,Pfister2016}).

Here we study this approach under the influence of noise and for parameters from the present ion trap setup in Innsbruck. Our figure of merit is the fidelity of the Bell pairs between repeater end nodes, which we require to be large enough to violate a Bell inequality. This is relevant for device-independent quantum key distribution (QKD) \cite{BellRMP}. We find that the scheme proposed in \cite{Sangouard09} does not allow one to beat direct transmission when taking current collective dephasing within the ion trap into account. We then propose a modified protocol where qubits are encoded in a decoherence free subspace (DFS) \cite{Zanardi1997,Lidar1998}, which is unaffected by collective dephasing, and show how the encoding and processing of logical Bell states can be done. Such a quantum repeater allows one to reach a distance of around 800 km between repeater end-nodes assuming state of the art parameters.

The outline of the paper is as follows. In Sec. \ref{system} we describe the system and present the relevant parameters. In Sec. \ref{noDFS} and Sec. \ref{DFS} we investigate the fidelities without and with a DFS, respectively. Sec. \ref{times} is devoted to the distribution times. In Sec. \ref{outlook} we outline how one could reach longer distances by either assuming higher fidelity operations or using logical entanglement purification, before we conclude in Sec. \ref{conclusions}.

\section{Protocol and system}
\label{system}
In this section we describe the direct transmission of quantum information through optical fibers, the quantum repeater protocol which we propose and the system which we are looking at, including the relevant noise processes and parameters.

\subsection{Direct transmission}
\label{direct}
The direct transmission of quantum information is done by sending an entangled pair of  photons at telecom wavelength (1550 nm) through an optical fiber. For short distances, noise can be neglected and the maximum reachable distance is limited by the (expected) time it takes a photon to arrive at the final destination. For a standard optical fiber the attenuation length is $L_{att}$ = 22 km, the speed of light in fiber is $c = 2 \cdot 10^{8}$ m/s and the transmission probability is $\eta_t = e^{-L_0 / L_{att}}$, where $L_0$ is the total length. We assume a detector efficiency of  $\eta_{d} = 0.9$ \cite{detector1,detector2} and a photon pair creation rate of 10 GHz, corresponding to state-of-the-art generation rates of entangled photons via parametric down conversion. One then finds that one can reach a distance of around 500 km with an average distribution time of around one second (see also Fig. \ref{rates}). Note that this protocol does not allow one to perform device-independent QKD due to its probabilistic nature.

\subsection{Proposed quantum repeater protocol}
The quantum repeater protocol which we propose here can be seen as an extension of \cite{Sangouard09}. In \cite{Sangouard09} entanglement is created between trapped ions in optical cavities and photons, which are sent through the channel. At intermediate locations these photons are measured in the Bell basis via linear optics in order to establish entanglement between neighboring quantum repeater nodes. Entanglement between the outermost quantum repeater nodes is achieved by performing (deterministic) Bell measurements within the ion traps at all intermediate quantum repeater stations.

Here we show how it is possible to encode the qubits into a DFS \cite{Zanardi1997,Lidar1998} that protects the quantum information from the effects of collective dephasing, an important source of decoherence in trapped ion systems \cite{Monz2011}. In addition we illustrate how the scheme can be extended to a scalable quantum repeater using logical entanglement purification and swapping.

\begin{widetext}

\begin{figure}[htb]
\centering
\includegraphics[scale=0.37]{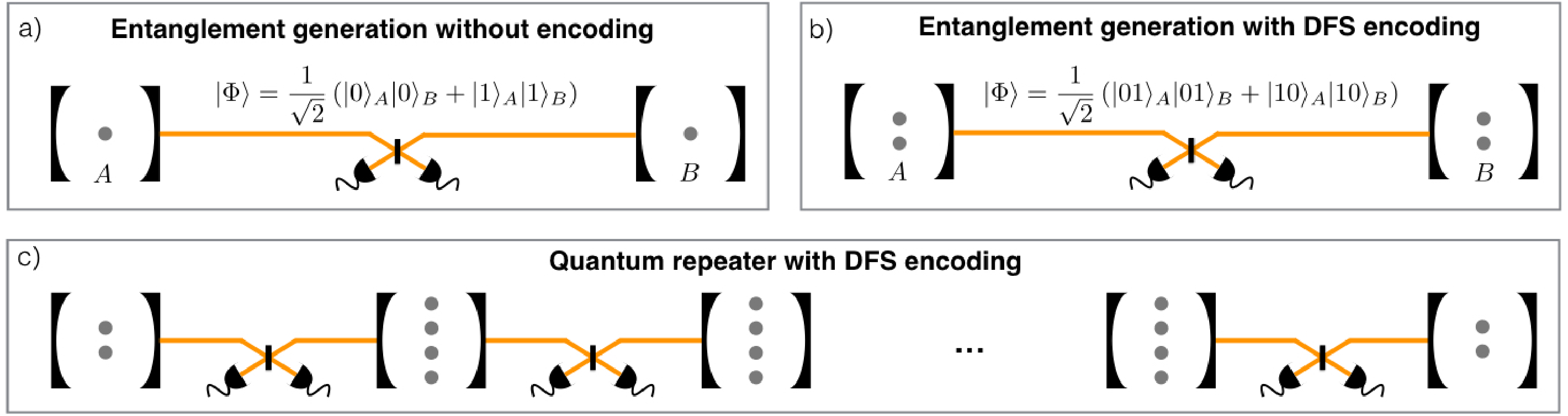}
\put(-185,10){$\dots$}
\caption{Illustration of the proposed quantum repeater scheme. At each node ion-photon entanglement is created. The photons are sent to intermediate locations, where a Bell measurement is performed using linear optics (a,b). The resulting Bell pairs between neighboring nodes (cavities) are called links. By performing Bell measurements on the ions at all intermediate nodes, entanglement is swapped over the entire distance (c). Collective dephasing within ion-traps \cite{Monz2011} is an important noise mechanism, which severely limits the achievable performance of such quantum repeaters without encoding (see panel a). This problem is mitigated by encoding quantum information in a decoherence free subspace (DFS), $\ket{0} \rightarrow \ket{01}$, $\ket{1} \rightarrow \ket{10}$, which renders the entangled links insensitive to collective dephasing. The connection of entangled links requires Bell measurements on the logical states, which can be achieved as described in Sec. \ref{logswap}.}
\label{illustration}
\end{figure}

\end{widetext}

To understand the timing for a quantum repeater and the role of decoherence, consider, for example, the elementary link between two nodes shown in Fig. 1a.  An ion at each node is entangled with a photon, and each photon propagates through optical fiber to an intermediate station.  For a 20 km link between nodes, each photon travels 10 km, corresponding to 50 ${\mu}$s in travel time through optical fiber.  At the station, a Bell measurement is performed, and the results are communicated back to each node over a classical channel, requiring an additional 50 ${\mu}$s. However, it is not possible to perform a two-photon Bell measurement with every entanglement attempt, as photons may be lost in the channel; it will be necessary to wait for several attempts before two photons arrive simultaneously.  On average, the wait time will be 100 ${\mu}$s / $P_{link}$, where $P_{link}$ is the probability to establish entanglement in a given attempt.  If this wait time is comparable to the decoherence time of an ion at each node, it means that our attempts to build up entanglement across many nodes will be undermined.  We address this problem by encoding the quantum state of single ion across two ions in a decoherence-free subspace (Fig. \ref{illustration}b).  As we build up entanglement across multiple links (Fig. \ref{illustration}c), the wait time remains the same, but we are no longer sensitive to decoherence.

For a more detailed discussion of timing for quantum repeaters see \cite{vanmeter2014,Jones2016}.

\subsection{System}
The physical system which we consider is the cavity quantum electrodynamics (CQED) setup in Innsbruck. It consists of an ion trap inside an optical cavity and allows one to create high fidelity ion-photon entanglement, state mapping between ions and photons, and to perform quantum logic between the trapped ions. For more details on ion-trap quantum networks and the role of cavities, see  \cite{Northup:2014aa,DuanRMP,Stute2012Appl}.
We use the following parameters:

\subsubsection{Ion-photon entanglement} 
Ion-photon entanglement can be created with 99.5\% fidelity \cite{Stute2012,Schindler2013}, which leads to ion-ion entanglement with 99\% fidelity. The state is assumed to be in Werner form, i.e. \be
 \rho=F\ket{\phi^+}\bra{\phi^+} +\tfrac{1-F}{3} \left( \ket{\phi^-}\bra{\phi^-} + \ket{\psi^+}\bra{\psi^+} + \ket{\psi^-}\bra{\psi^-} \right)
 \ee
 which can be enforced via depolarization \cite{Bennett96}. Werner states are a worst-case scenario and thus a conservative assumption. $\ket{\phi^{\pm}}$ and $\ket{\psi^{\pm}}$ denote the Bell states, i.e.,
 \be
 \ket{\phi^{\pm}} = \frac{1}{\sqrt{2}} \left( \ket{0}\ket{0} \pm \ket{1}\ket{1} \right)
 \ee
 and
 \be
 \ket{\psi^{\pm}} = \frac{1}{\sqrt{2}} \left( \ket{0}\ket{1} \pm \ket{1}\ket{0} \right).
 \ee
 
\subsubsection{Fiber and photon detector} 
As in \cite{Sangouard09} and in the direct transmission case above (Sec. \ref{direct}), we assume optical fiber transmission at 1550 nm wavelength with attenuation length $L_{att}$, and we use a detector efficiency $\eta_d = 0.9$.

\subsubsection{Probability of an ion emitting a single photon into the cavity output mode} 
In \cite{Stute2012}, the probability to generate an ion entangled with a photon and to couple that photon into fiber is 14\%, including losses in creating and collecting photons. In the same reference, it is argued that 99\% efficiency could be achieved in a future setup with state-of-the-art mirror losses and the choice of an output mirror with higher transmission. The scheme used in \cite{Stute2012} allows enhanced photon collection from a dipole transition at 854 nm, which has a branching ratio that would otherwise be unfavorable. This wavelength has great potential for efficient frequency conversion to the telecom band, as single-photon frequency conversion from red and near-infrared wavelengths to the telecom bands has been demonstrated with several tens of percent total efficiency, limited largely by passive optical losses \cite{Zaske2012,Albrecht2014,Pelc2011}. Note that we do not consider the detrimental effects of noise introduced by photon conversion: in principle this can be filtered out given the narrowband photons emitted by the cavity-ion photon source.

We thus choose a probability p = 0.35 for ion-photon entanglement including photon conversion, that is, we assume an efficiency of about 90\% for creation and collection of ion-photon entanglement in a future setup and a frequency conversion efficiency of about 40\%.  We note that these two processes have not yet been combined in an experimental setup.

\subsubsection{Photon repetition rate}
\label{photonrep}
In \cite{Casabone2013}, photon generation occurred every 440 $\mu$s, and this cycle was dominated by optical pumping.  The cycle was also interrupted every 4.4 ms to include 1.3 ms of Doppler cooling (77\% duty cycle). We estimate that it is experimentally realistic to generate photons every 20 $\mu$s, with cooling interleaved for 50 $\mu$s every ms (95\% duty cycle), based on improvements in optical pumping and cooling \cite{Lechner2016}.

In comparison, the repetition rate for the proposed quantum repeater protocol is set by twice the time a photon takes to reach the central station between two nodes at which a Bell measurement is performed.  For the distances of $\sim$ 500 km for which this protocol beats direct transmission, this corresponds to several hundred $\mu$s in optical fiber even for the highest number of links (that is, the shortest distance between links) that we analyze.  In this case, the estimated photon cycle time is much shorter than the transmission latency and can be neglected.

We also note that for the detectors of Sec. II.C.3, dead times are on the order of tens of ns and can be neglected.

\subsubsection{Noisy operations}
We model noisy operations by local depolarizing noise followed by the ideal map. For two-qubit operations we assume 0.5\% noise, for single-qubit operations 0.1\% noise.

The local depolarizing noise (LDN) is described by
\be
{\cal{D}}(p_g) \rho = p_g\rho + \frac{1-p_g}{4} \left(\rho+X\rho X + Y\rho Y + Z\rho Z \right)
\ee
where $X, Y, Z$ are the usual Pauli matrices and $1-p_g$ quantifies the level of noise. 
Noisy gates ${\hat U}$ are given by ${\hat U}\prod_a{\cal D}(p_{g})^{(a)} \rho$, i.e., single-qubit LDN with error parameter $p_g$ on all involved qubits,
followed by the ideal operation described by the superoperator ${\hat U}$ with ${\hat U}\rho = U \rho U^\dagger$, where $U$ is the unitary gate.

\subsubsection{Collective dephasing} 
Collective dephasing on $n$ qubits is described by the map
\be
\rho \rightarrow \int_{-\infty}^{\infty}e^{-i\theta \sum_{j=1}^{n}Z_j} \rho e^{i\theta \sum_{j=1}^{n}Z_j}p(\theta) d\theta 
\ee
with $p(\theta) = \frac{1}{\sqrt{2\pi}\sigma} e^{-\frac{\theta^2}{2\sigma^2}}$ and $\sigma = \frac{T}{\tau}$. $Z_j$ denotes the Pauli $Z$ operator on qubit $j$, $T$ is the time for which the ions have to be stored and $\tau$ is the coherence time, for which we assume $\tau = 10$ ms \cite{Monz2011,Schindler2013}. Collective dephasing can arise from fluctuations in control fields that couple to quantum systems globally. In ion traps, noise in the amplitude of magnetic fields that couple with equal strength to the closely-spaced ion can cause the ions' qubit transition energies to fluctuate in a correlated manner \cite{Monz2011}. In the Innsbruck setup, this effect currently limits the qubit coherence time to around 1 ms, and improvements to 10 ms are expected \cite{Schindler2013}. Further improvements would entail significant experimental overhead, e.g., ultra-stable magnetic field sources and an enclosure to shield the experiment from magnetic fields.

\subsubsection{Decoherence free subspace}
The benefits of using a DFS have been studied experimentally for $^{40}\rm{Ca}^{+}$ ions, the isotope we consider here, in \cite{Roos2004,Haffner2005}. It has been shown that entanglement can be preserved for up to 20 seconds and that on a time scale of one second, which is the relevant scale for the analysis here, the state fidelity did not decrease to within the measured precision \cite{Haffner2005}. The next limitation to the qubit lifetime arises from magnetic field gradients.  If necessary, when considering longer timescales, this limitation and others can be included in the value $p_g$, characterizing the quality of the ion-ion gates.

\section{Fidelities}
\label{fidelity}
In this section we investigate the achievable fidelities  for a trapped-ion-based quantum repeater, both when not using and when using a DFS. 

\subsection{Quantum repeater based on trapped ions without DFS}
\label{noDFS}
Here we study the original proposal \cite{Sangouard09} in the presence of noise. We assume the parameters given in Sec. \ref{system}. For the average distribution time we assume as a lower bound the time it takes to create entanglement in a single link. Consequently the corresponding fidelities represent bounds from above. We will show below that even with this optimistic assumption it is impossible to distribute entanglement over reasonable distances, i.e., distances of several hundred kilometers. The probability to establish entanglement within a link is given by \cite{Sangouard09}
\be
P_{link} = \frac{1}{2}p^2 \eta_t \eta_d^2,
\ee
and thus the average distribution time will be
\be
T = \tfrac{L_0}{c} \frac{1}{p^2 \eta_t \eta_d^2}
\ee
where $L_0$ is the length of an elementary link.

\begin{figure}[htb]
\centering
\includegraphics[scale=0.32]{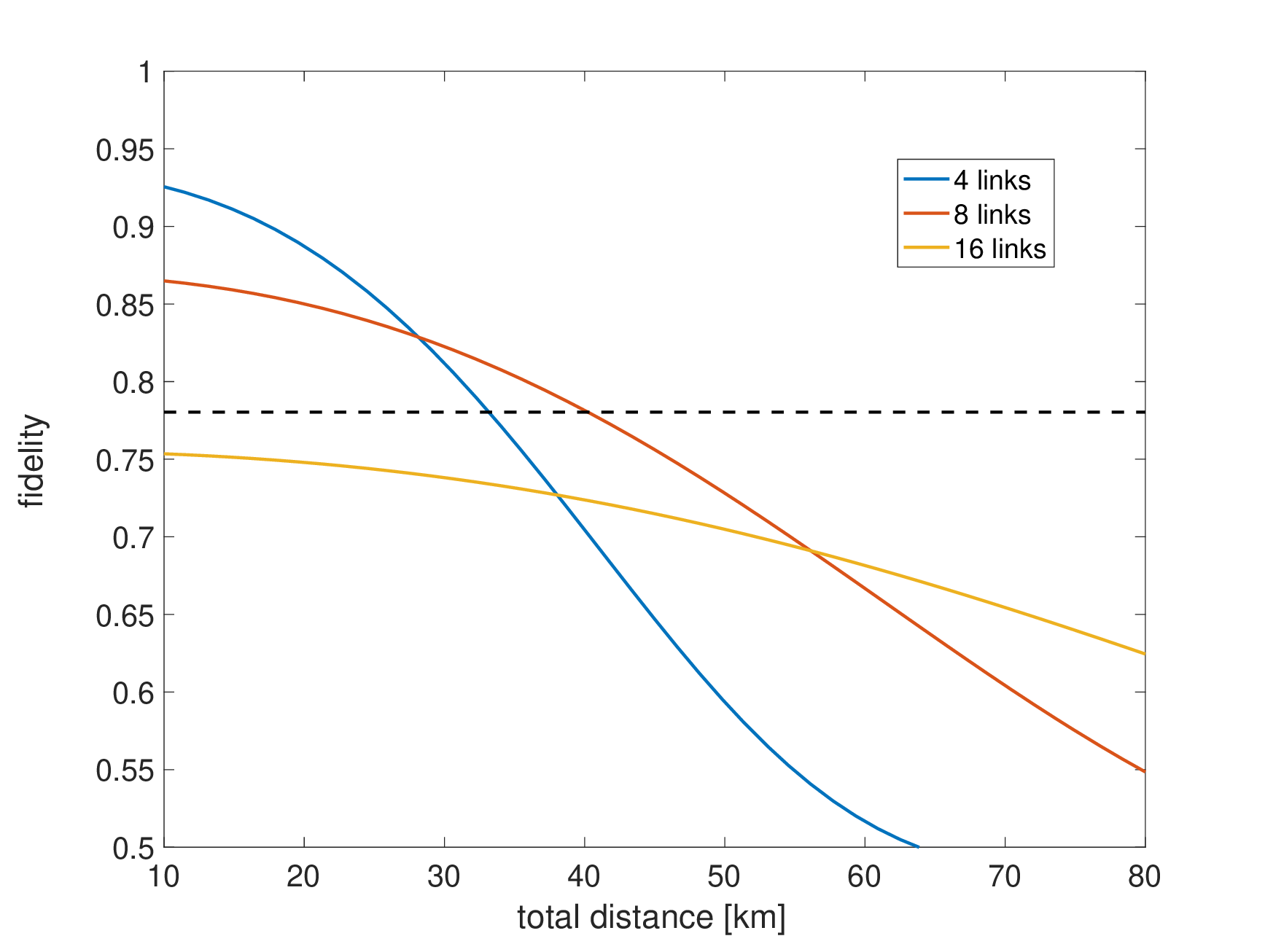}
\caption{Performance of a trapped-ion-based quantum repeater without a DFS. The fidelity is plotted as a function of (total) distance for four, eight, and 16 links. For the distribution time we assumed the time it takes to establish entanglement in one link, which is a lower bound (for more details see main text). In all cases it drops below 78\% (dashed line) before reaching 50 km, which implies that it does not outperform the direct transmission of photons. For more details see the main text.}
\label{IRfidelity}
\end{figure}

The resulting fidelities for ion-ion entanglement with respect to a maximally entangled state for four, eight, and 16 links are shown as a function of distance in Fig. \ref{IRfidelity}. The offset at zero distance is a result of the imperfect operations and non-unit fidelity of the Bell pairs in the elementary links. The drop as a function of distance is due to the fact that it takes longer to distribute a Bell pair over a longer distance, which means that the ions in each trap are exposed to collective dephasing for a longer time.  The fidelity drops below 78\%, which is the threshold required to violate a Bell inequality \cite{Bell,CHSH}, before reaching 50 km. Given the fact that the direct transmission of photons allows one to reach distances of around 500 km (see Fig. \ref{rates}) we conclude that a quantum repeater based on this architecture (and the noise parameters stated above) will not outperform the direct transmission of (telecom wavelength) photons through fibers. We note that in this best-case scenario, we have ignored the cycle time at each node for photon generation and re-initialization (see Sec. \ref{photonrep}). For link lengths of just a few kilometers, this cycle time is non-negligible but will only serve to degrade the repeater performance.  Thus, our conclusion that a quantum repeater will not outperform direct transmission holds, independent of the cycle time.

\subsection{Quantum repeater based on trapped ions with DFS}
\label{DFS}
In the previous subsection we have seen that (current) collective dephasing prevents us from beating direct transmission. The effect of collective dephasing can be compensated for by encoding the qubits in a DFS which is invariant under the noise. Notice that using the DFS is a passive form of protection. Unlike quantum error correction it does not require one to measure stabilizer operators and to perform subsequent operations conditioned on the outcomes of these measurements. Thus using the DFS only requires twice as many qubits (as compared to the case without a DFS), performing the encoding and logical swap operations instead of physical ones.

\subsubsection{Encoding}
The DFS which is invariant under collective dephasing is spanned by $\ket{0_L}=\ket{01}$ and $\ket{1_L}=\ket{10}$. An unknown state $\ket{\psi}=\alpha \ket{0} + \beta \ket{1}$ can be encoded by using an ancilla qubit in state $\ket{1}$ and applying a $CNOT$ gate with the ancilla qubit as target. The encoded state is then given by $\ket{\psi_L} = \alpha \ket{01} + \beta \ket{10}$.

\subsubsection{Logical entanglement swap operation}
\label{logswap}
The swapping of logical entanglement is performed by a logical Bell measurement. There are several possibilities to perform such a logical Bell measurement, two of which are listed below. The first version works probabilistically but allows one to detect errors. The second version is deterministic and includes no error detection.

\begin{figure}[htb]
\centering
\includegraphics[scale=0.4]{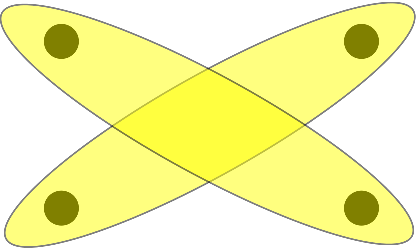}
\put(-90,45){$1$}
\put(-90,5){$3$}
\put(5,45){$2$}
\put(5,5){$4$}
\caption{Illustration of version 1 for a logical Bell measurement. The black dots represent the physical qubits. They are all in the same ion trap. Qubits 1 and 3 are part of one logical Bell pair, qubits 2 and 4 are part of the other logical Bell pair. Only entangling gates are shown (yellow ellipses).} 
\label{swap2}
\end{figure}
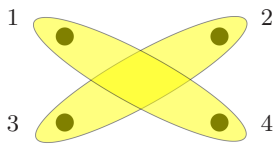

\paragraph*{Version 1 ---}
Here we describe a possibility which requires two entangling gates. Qubits 1 and 3 are on Alice's side, qubits 2 and 4 on Bob's side. The logical Bell measurement is achieved by performing the gates $CNOT^{3 \rightarrow 2}$ and $CNOT^{1 \rightarrow 4}$ and then measuring qubits 1 and 3 in the X basis, the other ones in the Z basis. An illustration is given in figure \ref{swap2}. The logical Bell state projection for given results of the single qubit measurements are given in table \ref{tableversion2}. Notice that in the case when one obtains measurement results that are not given in this table (which are all the results that one can obtain as long as the state is in the logical subspace), one discards the resulting state. This corresponds to an error detection scheme and leads to larger fidelities but also makes the protocol probabilistic.

\begin{table}
\caption{Logical Bell state projection for given results of the single qubit measurements for version 1.}
\vspace{0.3cm}
\centering
\begin{tabular}{| c | c | c | c | c |}
\hline
qubit $1$  & qubit $2$ & qubit $3$ & qubit $4$ & Bell state  \\ \hline
$+$ & $-$ &  $+$  & $-$ & $\ket{\phi^+_L}$\\
$-$ & $-$ &  $-$  & $-$ & $\ket{\phi^+_L}$\\
$+$ & $-$ &  $-$  & $-$ & $\ket{\phi^-_L}$\\
$-$ & $-$ &  $+$  & $-$ & $\ket{\phi^-_L}$\\
$+$ & $+$ &  $+$  & $+$ & $\ket{\psi^+_L}$\\
$-$ & $+$ &  $-$  & $+$ & $\ket{\psi^+_L}$\\
$+$ & $+$ &  $-$  & $+$ & $\ket{\psi^-_L}$\\
$-$ & $+$ &  $+$  & $+$ & $\ket{\psi^-_L}$\\
\hline
\end{tabular}
\label{tableversion2}
\end{table}

\begin{figure}[htb]
\centering
\includegraphics[scale=0.4]{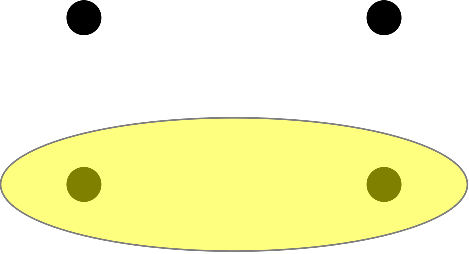}
\put(-100,45){$1$}
\put(-100,5){$3$}
\put(5,45){$2$}
\put(5,5){$4$}
\caption{Illustration of version 2 for a logical Bell measurement. The black dots represent the physical qubits. They are all in the same ion trap. Qubits 1 and 3 are part of one logical Bell pair, qubits 2 and 4 are part of the other logical Bell pair. Only entangling gates are shown (yellow ellipses).} 
\label{swap3}
\end{figure}
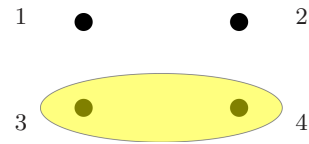

\paragraph*{Version 2 ---}
A version which requires only one entangling gate is given by the following protocol (the qubit labeling is the same as above): perform $CNOT^{3 \rightarrow 4}$ and measure qubits 1, 2 and 3 in the X basis, and qubit 4 in the Z basis. An illustration is given in figure \ref{swap3}. The logical Bell state projection for given results of the single qubit measurements are given in table \ref{tableversion3}.

\begin{table}
\caption{Logical Bell state projection for given results of the single qubit measurements for version 2.}
\vspace{0.3cm}
\centering
\begin{tabular}{| c | c | c | c | c |}
\hline
qubit $1$  & qubit $2$ & qubit $3$ & qubit $4$ & Bell state  \\ \hline
$+$ & $+$ &  $+$  & $+$ & $\ket{\phi^+_L}$\\
$+$ & $-$ &  $-$  & $+$ & $\ket{\phi^+_L}$\\
$- $& $+$ &  $-$  & $+$ & $\ket{\phi^+_L}$\\
$-$ & $-$ &  $+$  & $+$ & $\ket{\phi^+_L}$\\
$+$ & $+$ &  $-$  & $+$ & $\ket{\phi^-_L}$\\
$+$ & $-$ &  $+$  & $+$ & $\ket{\phi^-_L}$\\
$-$ & $+$ &  $-$  & $+$ & $\ket{\phi^-_L}$\\
$-$ & $-$ &  $+$  & $+$ & $\ket{\phi^-_L}$\\
$+$ & $+$ &  $+$ & $-$ & $\ket{\psi^+_L}$\\
$+$ &$ -$ &  $-$ & $-$ & $\ket{\psi^+_L}$\\
$- $& $+$ &  $-$ & $-$ & $\ket{\psi^+_L}$\\
$- $& $-$ &  $+$ & $-$ & $\ket{\psi^+_L}$\\
$+$ & $+$ &  $-$ & $-$ & $\ket{\psi^-_L}$\\
$+$ & $-$ &  $+$ & $-$ & $\ket{\psi^-_L}$\\
$-$ & $+$ &  $+$ & $-$ & $\ket{\psi^-_L}$\\
$-$ & $-$ &  $-$ & $-$ & $\ket{\psi^-_L}$\\
\hline
\end{tabular}
\label{tableversion3}
\end{table}

\subsubsection{Resulting fidelities}
The fidelities are computed under the assumption that all ions are stored for the entire distribution time (see SEC. \ref{times}). This ignores the possibility that ions at some nodes could be measured at an earlier time stage, namely when entanglement has been successfully generated in the adjacent links. Consequently these fidelities represent bounds from below. We compared the fidelities for various numbers of links using the different protocols described above. The version which uses error detection, version 1, performs better than version 2. The fidelities for the case when using version 1 are summarized in table \ref{tablefid1}, the ones for version 2 in table \ref{tablefid2}. These are the fidelities of the Bell pairs after the decoding, which is simply the inverse of the encoding circuit described above. There is a slight drop of the fidelity as a function of distance, despite the fact that the logical qubits are unaffected by collective dephasing due to the DFS. We attribute this to the imperfect (noisy) encoding, which leads to populations outside the logical subspace. Empirically we find that this drop is at most of the order of $0.5\%$ between distance zero and the distance which corresponds to a distribution time of a second. The fidelities which we present here are evaluated at a distance which corresponds to a distribution time of (at least) a second and thus represent a lower bound. The maximally reachable distance is limited by the distribution time (see Sec. \ref{times}). We analyzed the density matrices of the Bell pairs and found that they are diagonal in the Bell basis. Using depolarization \cite{Be96} one can bring them into Werner state form and thus describe them by
\be
\rho = {\cal{D}}_1(\tilde{p}) {\cal{D}}_2(\tilde{p}) \ket{\phi^+}\bra{\phi^+}.
\ee
In order to violate a Bell inequality \cite{Bell,CHSH} the fidelity needs to exceed $78.0\%$. This minimum fidelity puts a constraint on the maximum number of repeater links and thus on the maximally reachable distance for fixed distribution time.

\begin{table}
\caption{Fidelities for different numbers of links when using version 1. The maximal number of links such that the fidelity is still large enough to violate a Bell inequality is given by 10.}
\vspace{0.3cm}
\centering
\begin{tabular}{| c | c | c | c | c |}
\hline
$\#$ links & 4 & 8 & 10 & 11  \\ \hline
fidelity & $90.1\%$ & $82.3\%$ & $78.7\%$ & $77.0\%$ \\ \hline
\end{tabular}
\label{tablefid1}
\end{table}

\begin{table}
\caption{Fidelities for different numbers of links when using version 2. The maximal number of links such that the fidelity is still large enough to violate a Bell inequality is given by 7.}
\vspace{0.3cm}
\centering
\begin{tabular}{| c | c | c | c | c |}
\hline
$\#$ links & 4 & 6 & 7 & 8  \\ \hline
fidelity & $87.4\%$ & $81.9\%$ & $79.4\%$ & $76.9\%$ \\ \hline
\end{tabular}
\label{tablefid2}
\end{table}

\section{Distribution times}
\label{times}
The time it takes to establish an entangled pair is given by the time until there is entanglement in all links. An analytical approximation can be obtained by adding the times it takes until there is entanglement in at least one link given M links in which there is no entanglement ($M \in [1,N]$).  The entanglement waiting times in each segment are (independent) geometrically distributed random variables. Using the fact that the minimum of $n$ identical geometric random variables with success probability $p$ is itself geometric with success probability $1-(1-p)^n$ (see also \cite{Collins07}) one obtains for the (expected) distribution time $T$
\be
\label{Tdist}
T \approx \frac{L_0}{c} \sum_{n=1}^{N} \frac{1}{1-(1-P_{link})^n}
\ee
where $P_{link}$ is the success probability for creating entanglement in a single link, i.e. $P_{link}=\frac{1}{2}p^2 \eta_t \eta_d^2$, and $N$ is the total number of links. Eq. (\ref{Tdist}) is an accurate estimate of the distribution time as long as $NP_{link} \ll 1$, which is the case for the scenarios studied in this work (for $L_0 = 80 \text{km}$ we have $P_{link} = 0.0013$). 
We confirm the validity of eq. (\ref{Tdist}) by computing the distribution times numerically via an absorbing Markov chain. Here, a state $i$ is given by the number of links in which there is entanglement. The total number of states is given by $N+1$, the states $i=0...N-1$ are transient states, the state $N$ is absorbing. The transition matrix for this Markov chain is given by
\bea
    p_{ij}=\left\{
                \begin{array}{ll}
                  0 \quad\text{for} \quad i>j\\
                  {{N-i} \choose {j-i}} P_{link}^{j-i} (1-P_{link})^{N-j} \quad \text{for} \quad i \leq j.
                \end{array}
              \right.
\eea
We find very good agreement between the numerical simulation with the absorbing Markov chain and the analytical estimate eq. (\ref{Tdist}) with deviations of the order of one percent in the relevant regime, i.e., when the distribution time is approximately one second. Some concrete examples are shown in the Appendix \ref{App}. Thus we use eq. (\ref{Tdist}) for the remainder of this work.

The time for the classical communication (the measurement results from the logical Bell measurements need to be communicated) is bounded from above by $\tfrac{NL_0}{c}$.

When one uses version 1 for logical entanglement swapping one has to repeat the entanglement distribution on average $P^{-1}$ times, where $P$ is the probability that one obtains successful outcomes (i.e., the ones in table \ref{tableversion2}) for all entanglement swaps. 

\begin{figure}[htb]
\centering
\includegraphics[scale=0.17]{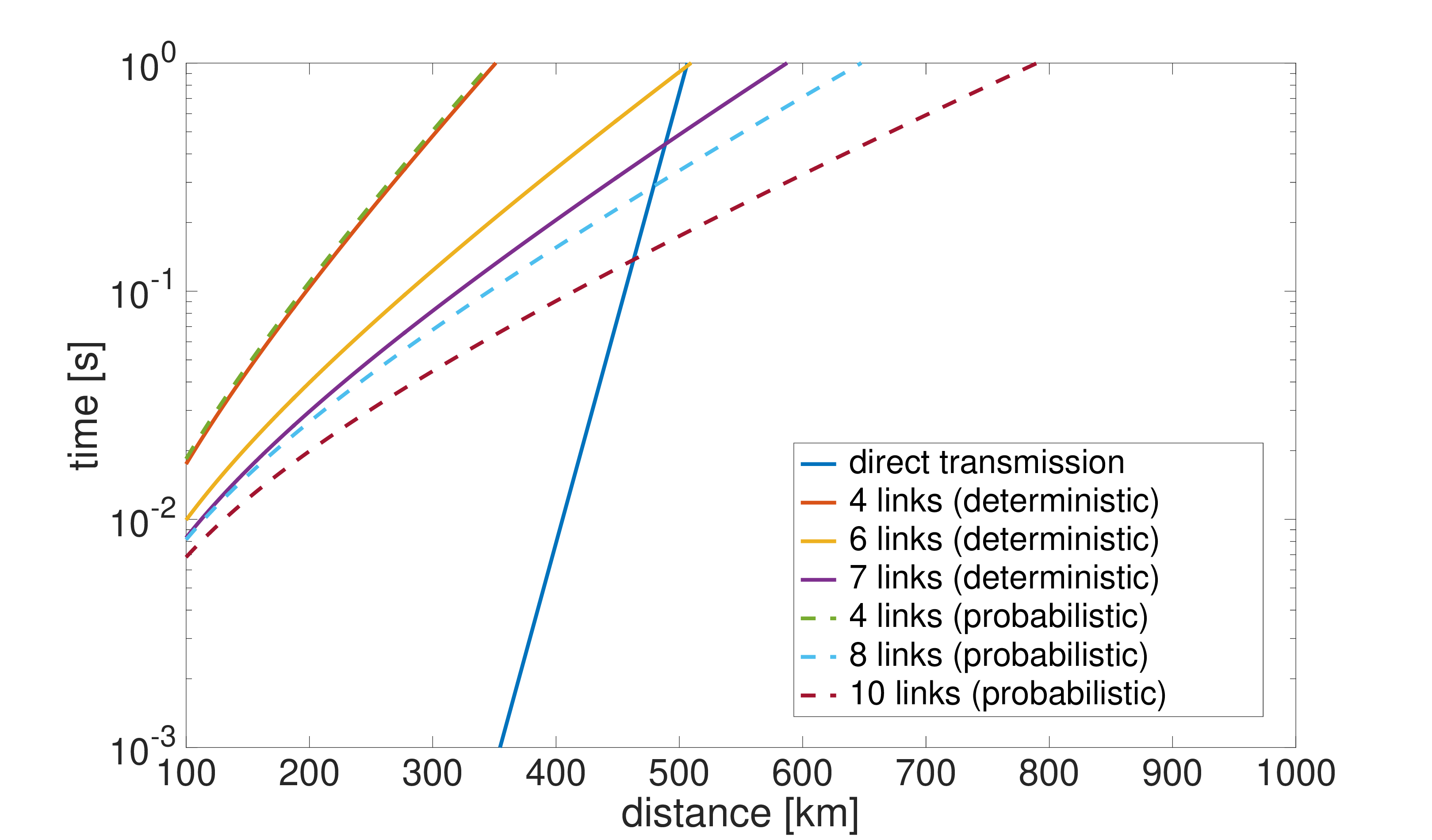}
\caption{Distribution time for various numbers of links and both versions for entanglement swapping and direct transmission for comparison. We assumed the following parameters: photon detector efficiency $\eta_{d} = 0.9$, probability of ion emitting a single photon $p = 0.35$, fiber attenuation length $L_{att}$ = 22 km and speed of light in fiber $c = 2 \cdot 10^{8}$ m/s. For the direct transmission we assumed a pair creation rate of 10 GHz.} 
\label{rates}
\end{figure}

The resulting distribution times are shown in Fig. \ref{rates}, where the direct transmission of a photon pair with a pair creation rate of 10 GHz is added for comparison. The parameters are those stated in Sec. \ref{system}.

We see that one can reach the largest distance, approximately 800 km, for a fixed distribution time of one second with version 1 for the logical entanglement swapping.

\section{Achieving longer distances}
\label{outlook}
There are different strategies which would allow us to reach longer distances. Two possibilities are sketched below, both of which require more advanced technology.

\subsection{Higher fidelity operations}
One possibility is to increase the fidelity of the ion-photon entanglement and the gates. The resulting fidelities, when assuming ion-ion entanglement with 99.9\% fidelity and single-(two) qubit gates with 0.01 (0.1)\% noise, are provided in tables \ref{fidoutlook1} and \ref{fidoutlook2} for version 1 and 2 of the swapping operations, respectively. The assumed noise parameters are beyond what has been demonstrated in the current setup, but it should be noted that gates with such low noise have already been achieved in other ion trap setups \cite{Ballance15}.

\begin{figure}[htb]
\centering
\includegraphics[scale=0.17]{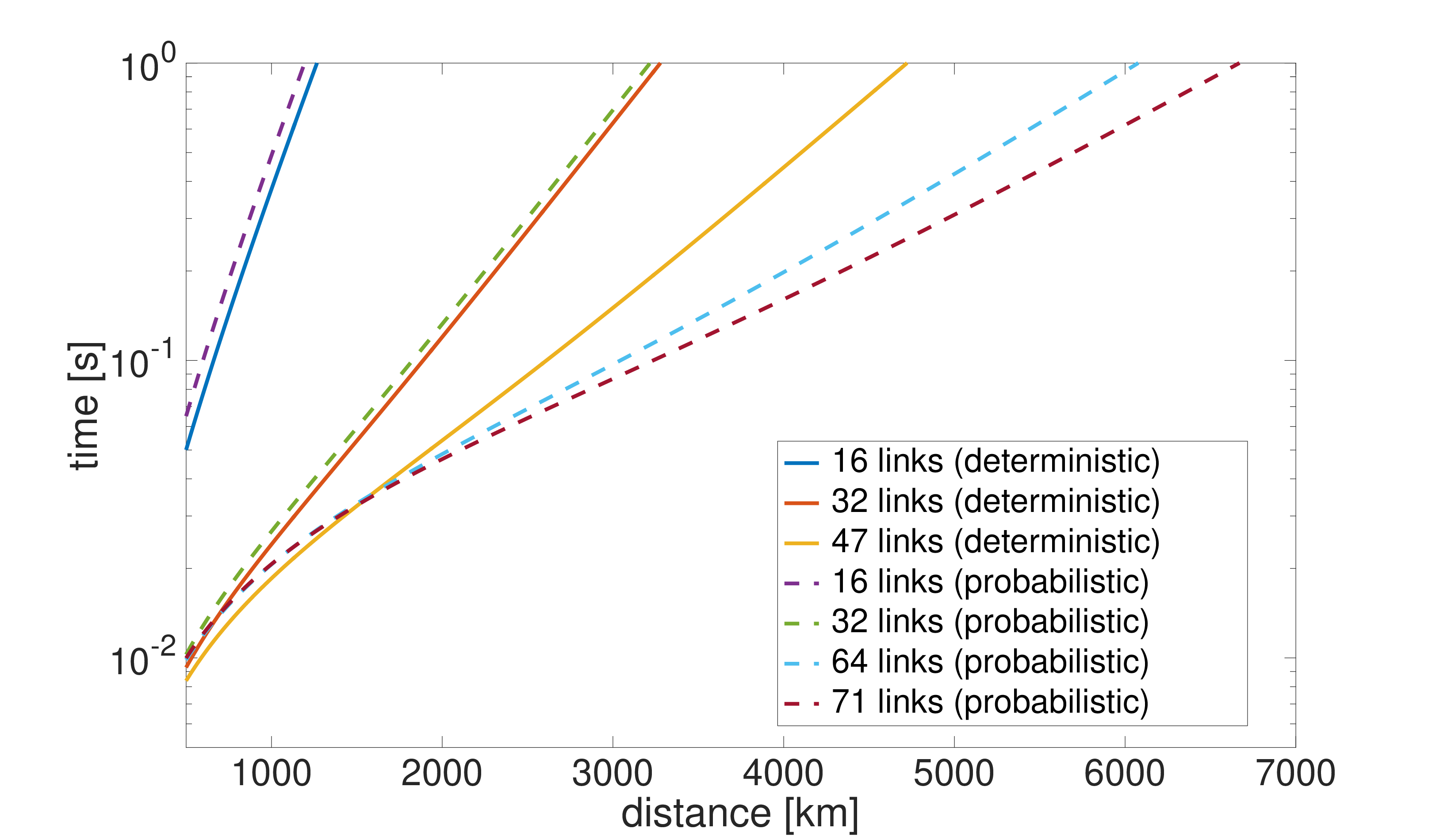}
\caption{Distribution time for various numbers of links and both versions of entanglement swapping assuming that the probability of an ion emitting a single photon followed by successful frequency conversion is $p=0.75$. The other parameters are: photon detector efficiency $\eta_{d} = 0.9$, fiber attenuation length $L_{att}$ = 22 km and speed of light in fiber $c = 2 \cdot 10^{8}$ m/s.} 
\label{timesoutlook}
\end{figure}

The corresponding distribution times are plotted in Fig. \ref{timesoutlook}, assuming that the probability of an ion emitting a single photon followed by successful frequency conversion is given by $p=0.75$. Ion-cavity setups and photon conversion devices are not available now with these efficiencies, but such probabilities are reasonable assuming state-of-the-art cavities (see the discussion in the Methods section of Ref. \cite{Stute2012}) and the fact that there is no fundamental limit to photon conversion efficiency.

In summary, we see that the quantum repeater based on trapped ions and using a DFS as described above is capable of extending quantum communication to distances of several thousand kilometers assuming futuristic but reasonable parameters.

\begin{table}[h]
\caption{Fidelities for different numbers of links when assuming that the fidelity of ion-ion entanglement is given by 99.9\% and 0.01 (0.1)\% noise for single-(two) qubit gates. Here we use version 1 for the logical entanglement swapping.}
\vspace{0.3cm}
\centering
\begin{tabular}{| c | c | c | c | c | c | c |}
\hline
$\#$ links & 16 & 32 & 64 & 70 & 71 & 72  \\ \hline
fidelity & $96.9\%$ & $ 89.0\% $ & $79.9\%$ & $78.3\%$ & $78.1\%$ & $77.8\%$\\
\hline
\end{tabular}
\label{fidoutlook1}
\end{table}

\begin{table}[h]
\caption{Fidelities for different numbers of links when assuming that the fidelity of ion-ion entanglement is given by 99.9\% and 0.01 (0.1)\% noise for single-(two) qubit gates. Here we use version 2 for the logical entanglement swapping.}
\vspace{0.3cm}
\centering
\begin{tabular}{| c | c | c | c | c |}
\hline
$\#$ links & 16 & 32 & 47 & 48  \\ \hline
fidelity & $91.7\%$ & $ 84.4\% $ & $78.2\%$ & $77.9\%$\\
\hline
\end{tabular}
\label{fidoutlook2}
\end{table}

\subsection{Full quantum repeater with entanglement purification  and DFS encoding}
A second possibility to increase the maximally reachable distance is to include entanglement purification \cite{Bennett96,Bennett96a,Deutsch96}, so that the scheme becomes scalable \cite{Briegel98}. The idea is the following: the Bell pairs in each link are encoded into a DFS in order to protect them from collective dephasing. Subsequently one performs entanglement purification and swapping of logical Bell pairs in a nested way \cite{Briegel98}. This requires significantly larger quantum memories at each node \cite{Briegel98,Duer1999}. For related work focusing on a setup with cold neutral atoms see \cite{Klein2006,Dorner2008}.

In order to implement purification one needs to be able to perform Clifford gates and Pauli measurements. Below we provide a set of generators of logical Clifford gates (qubits 1 and 2 (3 and 4) encode the first (second) logical qubit) for the encoding $\ket{0_L}=\ket{01}$ and $\ket{1_L}=\ket{10}$:
\bea
S_L &=& S^{1}\\ \nonumber
H_L &=& \left[(HSHZ)\otimes(HSH)\right] CNOT^{1\rightarrow2} \left[(HSX)\otimes X\right] \\ \nonumber
CZ_L &=& CZ^{1\rightarrow3},
\eea
where $S = \text{diag}(1, i)$ (in the computational basis) is a rotation around the $Z$-axis by an angle $\tfrac{\pi}{2}$, $H$ is the Hadamard gate and $CZ = \text{diag}(1, 1, 1, -1)$ denotes the controlled phase gate.
A measurement in the logical computational basis can be performed by measuring the first physical qubit in the computational basis. Notice that any Clifford circuit can be realized by combining these generating operations. However, for particular protocols a direct implementation of the required logical unitary operation using fewer two-qubit gates might be possible.

This allows one to implement one-way and two-way entanglement purification protocols at the logical level. In particular, one can implement the recurrence protocol of \cite{Deutsch96} that is used in the scalable repeater scheme of \cite{Briegel98}.  In this case, entanglement between logical states is established between neighboring nodes, where two (or more) basic links are connected. The DFS encoding protects the system against memory errors. Before the entanglement drops below a certain threshold value, entanglement purification at the logical level is used to re-establish long-distance pairs with increased fidelity.
We remark that also schemes based on error correction, i.e. transmission of encoded information with a DFS encoding at the lowest level, can be implemented in this way when using CSS codes, as also in this case all required operations are of Clifford type.

\section{Conclusions}
\label{conclusions}
We analyzed the quantum repeater based on trapped ions \cite{Sangouard09} for realistic parameters for qubit coherence, logic gate operations, ion-photon entanglement, and photon collection and detection efficiencies.  We have found that the repeater's performance is severely limited by collective dephasing within the ion traps. We then proposed a modified protocol where the qubits are encoded in a decoherence free subspace and showed that this allows one to reach distances of approximately 800 km for state of the art parameters. By considering parameters that have not been achieved yet, but can realistically be expected to be obtainable in the trapped-ion platform, we found that the reachable distance can be extended by almost an order of magnitude to several thousand kilometers. This allows one to reach intercontinental distances, even without entanglement purification, with fidelities that are even high enough to violate Bell inequalities. In future work, it will be important to extend our analysis to include the limitations imposed by other sources of error, e.g., photon indistinguishability.

The usage of a decoherence free subspace as a lowest layer of protection against dominant noise sources is not restricted to ion trap set-ups. One can use a similar approach in other systems, e.g. NV-centers (see \cite{Reiserer16}). Finally, we remark that higher rates can in principle be reached using an approach based on the transmission of encoded information (see \cite{Knill96,Zw14,Mur14}). However, this comes at the price of a large overhead in required resources, on the one hand due to small repeater spacing (of the order of 1 km) and on the other hand due to the large number of ions to be processed at each node because one uses a (concatenated) error correction code. To compare the performance and resource requirements of such a scheme to our approach for the same set-up would be interesting.

\section*{Acknowledgements}
This work was supported by the Swiss National Science Foundation (SNSF) through Grant number PP00P2-150579, the Army Research Laboratory Center for Distributed Quantum Information via the project SciNet, the EU via the integrated project SIQS, and the Austrian Science Fund (FWF): P28000 - N27, F4019 and V252. B. P. Lanyon acknowledges the 2015 Austrian START prize, with project number Y 849-N20.

\section{Appendix}
\label{App}
Here we provide some concrete examples for the distribution times when calculated with the analytical approximation eq. (\ref{Tdist}) and the absorbing Markov chain. They are summarized for seven links in table \ref{comp}. The numbers agree very well, and similar behavior is found for different numbers of links.

\begin{table}[h]
\caption{Distribution times for a quantum repeater with seven links when using the analytical approximation and the Markov chain approach.}
\vspace{0.3cm}
\centering
\begin{tabular}{| c | c | c | c | c | c |}
\hline
distance in km & 400 & 450 & 500 & 550 & 600  \\ \hline
time in s (analytical) & $0.2052$ & $0.3173$ & $0.4856$ & $0.7366$ & $1.1092$\\ \hline
time in s (Markov) & $0.2003$ & $0.3119$ & $0.4795$ & $0.7299$ & $1.1019$\\
\hline
\end{tabular}
\label{comp}
\end{table}

\bibliographystyle{apsrev4-1}
\bibliography{DFSbib}

\begin{thebibliography}{54}%
\makeatletter
\providecommand \@ifxundefined [1]{%
 \@ifx{#1\undefined}
}%
\providecommand \@ifnum [1]{%
 \ifnum #1\expandafter \@firstoftwo
 \else \expandafter \@secondoftwo
 \fi
}%
\providecommand \@ifx [1]{%
 \ifx #1\expandafter \@firstoftwo
 \else \expandafter \@secondoftwo
 \fi
}%
\providecommand \natexlab [1]{#1}%
\providecommand \enquote  [1]{``#1''}%
\providecommand \bibnamefont  [1]{#1}%
\providecommand \bibfnamefont [1]{#1}%
\providecommand \citenamefont [1]{#1}%
\providecommand \href@noop [0]{\@secondoftwo}%
\providecommand \href [0]{\begingroup \@sanitize@url \@href}%
\providecommand \@href[1]{\@@startlink{#1}\@@href}%
\providecommand \@@href[1]{\endgroup#1\@@endlink}%
\providecommand \@sanitize@url [0]{\catcode `\\12\catcode `\$12\catcode
  `\&12\catcode `\#12\catcode `\^12\catcode `\_12\catcode `\%12\relax}%
\providecommand \@@startlink[1]{}%
\providecommand \@@endlink[0]{}%
\providecommand \url  [0]{\begingroup\@sanitize@url \@url }%
\providecommand \@url [1]{\endgroup\@href {#1}{\urlprefix }}%
\providecommand \urlprefix  [0]{URL }%
\providecommand \Eprint [0]{\href }%
\providecommand \doibase [0]{http://dx.doi.org/}%
\providecommand \selectlanguage [0]{\@gobble}%
\providecommand \bibinfo  [0]{\@secondoftwo}%
\providecommand \bibfield  [0]{\@secondoftwo}%
\providecommand \translation [1]{[#1]}%
\providecommand \BibitemOpen [0]{}%
\providecommand \bibitemStop [0]{}%
\providecommand \bibitemNoStop [0]{.\EOS\space}%
\providecommand \EOS [0]{\spacefactor3000\relax}%
\providecommand \BibitemShut  [1]{\csname bibitem#1\endcsname}%
\let\auto@bib@innerbib\@empty
\bibitem [{\citenamefont {Bennett}\ and\ \citenamefont
  {Brassard}(1984)}]{BB84}%
  \BibitemOpen
  \bibfield  {author} {\bibinfo {author} {\bibfnamefont {C.~H.}\ \bibnamefont
  {Bennett}}\ and\ \bibinfo {author} {\bibfnamefont {G.}~\bibnamefont
  {Brassard}},\ }in\ \href@noop {} {\emph {\bibinfo {booktitle} {Proceedings of
  IEEE International Conference on Computers, Systems and Signal Processing}}}\
  (\bibinfo  {publisher} {IEEE Computer Society},\ \bibinfo {address}
  {Bangalore, India},\ \bibinfo {year} {1984})\ pp.\ \bibinfo {pages}
  {175--179}\BibitemShut {NoStop}%
\bibitem [{\citenamefont {Ekert}(1991)}]{E91}%
  \BibitemOpen
  \bibfield  {author} {\bibinfo {author} {\bibfnamefont {A.~K.}\ \bibnamefont
  {Ekert}},\ }\href {\doibase 10.1103/PhysRevLett.67.661} {\bibfield  {journal}
  {\bibinfo  {journal} {Phys. Rev. Lett.}\ }\textbf {\bibinfo {volume} {67}},\
  \bibinfo {pages} {661} (\bibinfo {year} {1991})}\BibitemShut {NoStop}%
\bibitem [{\citenamefont {Scarani}\ \emph {et~al.}(2009)\citenamefont
  {Scarani}, \citenamefont {Bechmann-Pasquinucci}, \citenamefont {Cerf},
  \citenamefont {Du\ifmmode~\check{s}\else \v{s}\fi{}ek}, \citenamefont
  {L\"utkenhaus},\ and\ \citenamefont {Peev}}]{Scarani09}%
  \BibitemOpen
  \bibfield  {author} {\bibinfo {author} {\bibfnamefont {V.}~\bibnamefont
  {Scarani}}, \bibinfo {author} {\bibfnamefont {H.}~\bibnamefont
  {Bechmann-Pasquinucci}}, \bibinfo {author} {\bibfnamefont {N.~J.}\
  \bibnamefont {Cerf}}, \bibinfo {author} {\bibfnamefont {M.}~\bibnamefont
  {Du\ifmmode~\check{s}\else \v{s}\fi{}ek}}, \bibinfo {author} {\bibfnamefont
  {N.}~\bibnamefont {L\"utkenhaus}}, \ and\ \bibinfo {author} {\bibfnamefont
  {M.}~\bibnamefont {Peev}},\ }\href {\doibase 10.1103/RevModPhys.81.1301}
  {\bibfield  {journal} {\bibinfo  {journal} {Rev. Mod. Phys.}\ }\textbf
  {\bibinfo {volume} {81}},\ \bibinfo {pages} {1301} (\bibinfo {year}
  {2009})}\BibitemShut {NoStop}%
\bibitem [{\citenamefont {Meter}\ and\ \citenamefont {Oskin}(2006)}]{Meter06}%
  \BibitemOpen
  \bibfield  {author} {\bibinfo {author} {\bibfnamefont {R.~V.}\ \bibnamefont
  {Meter}}\ and\ \bibinfo {author} {\bibfnamefont {M.}~\bibnamefont {Oskin}},\
  }\href {\doibase 10.1145/1126257.1126259} {\bibfield  {journal} {\bibinfo
  {journal} {J. Emerg. Technol. Comput. Syst.}\ }\textbf {\bibinfo {volume}
  {2}},\ \bibinfo {pages} {31} (\bibinfo {year} {2006})}\BibitemShut {NoStop}%
\bibitem [{\citenamefont {Meter}\ \emph {et~al.}(2008)\citenamefont {Meter},
  \citenamefont {Munro}, \citenamefont {Nemoto},\ and\ \citenamefont
  {Itoh}}]{Meter08}%
  \BibitemOpen
  \bibfield  {author} {\bibinfo {author} {\bibfnamefont {R.~V.}\ \bibnamefont
  {Meter}}, \bibinfo {author} {\bibfnamefont {W.~J.}\ \bibnamefont {Munro}},
  \bibinfo {author} {\bibfnamefont {K.}~\bibnamefont {Nemoto}}, \ and\ \bibinfo
  {author} {\bibfnamefont {K.~M.}\ \bibnamefont {Itoh}},\ }\href {\doibase
  10.1145/1324177.1324179} {\bibfield  {journal} {\bibinfo  {journal} {J.
  Emerg. Technol. Comput. Syst.}\ }\textbf {\bibinfo {volume} {3}},\ \bibinfo
  {pages} {2:1} (\bibinfo {year} {2008})}\BibitemShut {NoStop}%
\bibitem [{\citenamefont {Beals}\ \emph {et~al.}(2013)\citenamefont {Beals},
  \citenamefont {Brierley}, \citenamefont {Gray}, \citenamefont {Harrow},
  \citenamefont {Kutin}, \citenamefont {Linden}, \citenamefont {Shepherd},\
  and\ \citenamefont {Stather}}]{Beals20120686}%
  \BibitemOpen
  \bibfield  {author} {\bibinfo {author} {\bibfnamefont {R.}~\bibnamefont
  {Beals}}, \bibinfo {author} {\bibfnamefont {S.}~\bibnamefont {Brierley}},
  \bibinfo {author} {\bibfnamefont {O.}~\bibnamefont {Gray}}, \bibinfo {author}
  {\bibfnamefont {A.~W.}\ \bibnamefont {Harrow}}, \bibinfo {author}
  {\bibfnamefont {S.}~\bibnamefont {Kutin}}, \bibinfo {author} {\bibfnamefont
  {N.}~\bibnamefont {Linden}}, \bibinfo {author} {\bibfnamefont
  {D.}~\bibnamefont {Shepherd}}, \ and\ \bibinfo {author} {\bibfnamefont
  {M.}~\bibnamefont {Stather}},\ }\href
  {http://rspa.royalsocietypublishing.org/content/469/2153/20120686} {\bibfield
   {journal} {\bibinfo  {journal} {Proceedings of the Royal Society of London
  A: Mathematical, Physical and Engineering Sciences}\ }\textbf {\bibinfo
  {volume} {469}} (\bibinfo {year} {2013})}\BibitemShut {NoStop}%
\bibitem [{\citenamefont {Bartlett}\ \emph {et~al.}(2007)\citenamefont
  {Bartlett}, \citenamefont {Rudolph},\ and\ \citenamefont
  {Spekkens}}]{BartlettRMP}%
  \BibitemOpen
  \bibfield  {author} {\bibinfo {author} {\bibfnamefont {S.~D.}\ \bibnamefont
  {Bartlett}}, \bibinfo {author} {\bibfnamefont {T.}~\bibnamefont {Rudolph}}, \
  and\ \bibinfo {author} {\bibfnamefont {R.~W.}\ \bibnamefont {Spekkens}},\
  }\href {\doibase 10.1103/RevModPhys.79.555} {\bibfield  {journal} {\bibinfo
  {journal} {Rev. Mod. Phys.}\ }\textbf {\bibinfo {volume} {79}},\ \bibinfo
  {pages} {555} (\bibinfo {year} {2007})}\BibitemShut {NoStop}%
\bibitem [{\citenamefont {Islam}\ \emph {et~al.}(2014)\citenamefont {Islam},
  \citenamefont {Magnin}, \citenamefont {Sorg},\ and\ \citenamefont
  {Wehner}}]{Islam2014}%
  \BibitemOpen
  \bibfield  {author} {\bibinfo {author} {\bibfnamefont {T.}~\bibnamefont
  {Islam}}, \bibinfo {author} {\bibfnamefont {L.}~\bibnamefont {Magnin}},
  \bibinfo {author} {\bibfnamefont {B.}~\bibnamefont {Sorg}}, \ and\ \bibinfo
  {author} {\bibfnamefont {S.}~\bibnamefont {Wehner}},\ }\href
  {http://stacks.iop.org/1367-2630/16/i=6/a=063040} {\bibfield  {journal}
  {\bibinfo  {journal} {New Journal of Physics}\ }\textbf {\bibinfo {volume}
  {16}},\ \bibinfo {pages} {063040} (\bibinfo {year} {2014})}\BibitemShut
  {NoStop}%
\bibitem [{\citenamefont {Gottesman}\ \emph {et~al.}(2012)\citenamefont
  {Gottesman}, \citenamefont {Jennewein},\ and\ \citenamefont
  {Croke}}]{Gottesman2012}%
  \BibitemOpen
  \bibfield  {author} {\bibinfo {author} {\bibfnamefont {D.}~\bibnamefont
  {Gottesman}}, \bibinfo {author} {\bibfnamefont {T.}~\bibnamefont
  {Jennewein}}, \ and\ \bibinfo {author} {\bibfnamefont {S.}~\bibnamefont
  {Croke}},\ }\href {\doibase 10.1103/PhysRevLett.109.070503} {\bibfield
  {journal} {\bibinfo  {journal} {Phys. Rev. Lett.}\ }\textbf {\bibinfo
  {volume} {109}},\ \bibinfo {pages} {070503} (\bibinfo {year}
  {2012})}\BibitemShut {NoStop}%
\bibitem [{\citenamefont {Broadbent}\ \emph {et~al.}(2009)\citenamefont
  {Broadbent}, \citenamefont {Fitzsimons},\ and\ \citenamefont
  {Kashefi}}]{Broadbent2009}%
  \BibitemOpen
  \bibfield  {author} {\bibinfo {author} {\bibfnamefont {A.}~\bibnamefont
  {Broadbent}}, \bibinfo {author} {\bibfnamefont {J.}~\bibnamefont
  {Fitzsimons}}, \ and\ \bibinfo {author} {\bibfnamefont {E.}~\bibnamefont
  {Kashefi}},\ }in\ \href {\doibase 10.1109/FOCS.2009.36} {\emph {\bibinfo
  {booktitle} {2009 50th Annual IEEE Symposium on Foundations of Computer
  Science}}}\ (\bibinfo {year} {2009})\ pp.\ \bibinfo {pages}
  {517--526}\BibitemShut {NoStop}%
\bibitem [{\citenamefont {{H. J. Kimble}}(2008)}]{Kimble08}%
  \BibitemOpen
  \bibfield  {author} {\bibinfo {author} {\bibnamefont {{H. J. Kimble}}},\
  }\href {\doibase 10.1038/nature07127} {\bibfield  {journal} {\bibinfo
  {journal} {{Nature}}\ }\textbf {\bibinfo {volume} {{453}}},\ \bibinfo {pages}
  {1023} (\bibinfo {year} {{2008}})}\BibitemShut {NoStop}%
\bibitem [{\citenamefont {Wootters}\ and\ \citenamefont
  {Zurek}(1982)}]{Wootters82}%
  \BibitemOpen
  \bibfield  {author} {\bibinfo {author} {\bibfnamefont {W.}~\bibnamefont
  {Wootters}}\ and\ \bibinfo {author} {\bibfnamefont {W.}~\bibnamefont
  {Zurek}},\ }\href@noop {} {\bibfield  {journal} {\bibinfo  {journal}
  {Nature}\ }\textbf {\bibinfo {volume} {299}},\ \bibinfo {pages} {802}
  (\bibinfo {year} {1982})}\BibitemShut {NoStop}%
\bibitem [{\citenamefont {Knill}\ and\ \citenamefont
  {Laflamme}(1996)}]{Knill96}%
  \BibitemOpen
  \bibfield  {author} {\bibinfo {author} {\bibfnamefont {E.}~\bibnamefont
  {Knill}}\ and\ \bibinfo {author} {\bibfnamefont {R.}~\bibnamefont
  {Laflamme}},\ }\href@noop {} {\enquote {\bibinfo {title} {Concatenated
  {Q}uantum {C}odes},}\ } (\bibinfo {year} {1996}),\ \Eprint
  {http://arxiv.org/abs/arXiv:quant-ph/9608012} {arXiv:quant-ph/9608012}
  \BibitemShut {NoStop}%
\bibitem [{\citenamefont {Briegel}\ \emph {et~al.}(1998)\citenamefont
  {Briegel}, \citenamefont {D{\"u}r}, \citenamefont {Cirac},\ and\
  \citenamefont {Zoller}}]{Briegel98}%
  \BibitemOpen
  \bibfield  {author} {\bibinfo {author} {\bibfnamefont {H.-J.}\ \bibnamefont
  {Briegel}}, \bibinfo {author} {\bibfnamefont {W.}~\bibnamefont {D{\"u}r}},
  \bibinfo {author} {\bibfnamefont {J.~I.}\ \bibnamefont {Cirac}}, \ and\
  \bibinfo {author} {\bibfnamefont {P.}~\bibnamefont {Zoller}},\ }\href
  {\doibase 10.1103/PhysRevLett.81.5932} {\bibfield  {journal} {\bibinfo
  {journal} {Phys. Rev. Lett.}\ }\textbf {\bibinfo {volume} {81}},\ \bibinfo
  {pages} {5932} (\bibinfo {year} {1998})}\BibitemShut {NoStop}%
\bibitem [{\citenamefont {D\"ur}\ \emph {et~al.}(1999)\citenamefont {D\"ur},
  \citenamefont {Briegel}, \citenamefont {Cirac},\ and\ \citenamefont
  {Zoller}}]{Duer1999}%
  \BibitemOpen
  \bibfield  {author} {\bibinfo {author} {\bibfnamefont {W.}~\bibnamefont
  {D\"ur}}, \bibinfo {author} {\bibfnamefont {H.-J.}\ \bibnamefont {Briegel}},
  \bibinfo {author} {\bibfnamefont {J.~I.}\ \bibnamefont {Cirac}}, \ and\
  \bibinfo {author} {\bibfnamefont {P.}~\bibnamefont {Zoller}},\ }\href
  {\doibase 10.1103/PhysRevA.59.169} {\bibfield  {journal} {\bibinfo  {journal}
  {Phys. Rev. A}\ }\textbf {\bibinfo {volume} {59}},\ \bibinfo {pages} {169}
  (\bibinfo {year} {1999})}\BibitemShut {NoStop}%
\bibitem [{\citenamefont {Muralidharan}\ \emph {et~al.}(2016)\citenamefont
  {Muralidharan}, \citenamefont {Li}, \citenamefont {Kim}, \citenamefont
  {L{\"u}tkenhaus}, \citenamefont {Lukin},\ and\ \citenamefont {Jiang}}]{Mu16}%
  \BibitemOpen
  \bibfield  {author} {\bibinfo {author} {\bibfnamefont {S.}~\bibnamefont
  {Muralidharan}}, \bibinfo {author} {\bibfnamefont {L.}~\bibnamefont {Li}},
  \bibinfo {author} {\bibfnamefont {J.}~\bibnamefont {Kim}}, \bibinfo {author}
  {\bibfnamefont {N.}~\bibnamefont {L{\"u}tkenhaus}}, \bibinfo {author}
  {\bibfnamefont {M.~D.}\ \bibnamefont {Lukin}}, \ and\ \bibinfo {author}
  {\bibfnamefont {L.}~\bibnamefont {Jiang}},\ }\href
  {http://dx.doi.org/10.1038/srep20463} {\bibfield  {journal} {\bibinfo
  {journal} {Scientific Reports}\ }\textbf {\bibinfo {volume} {6}},\ \bibinfo
  {pages} {20463 EP } (\bibinfo {year} {2016})}\BibitemShut {NoStop}%
\bibitem [{Note1()}]{Note1}%
  \BibitemOpen
  \bibinfo {note} {A quantum repeater is scalable if one can create a Bell pair
  with a fidelity exceeding a predefined value over arbitrarily long distances
  with at most polynomially growing resources}\BibitemShut {NoStop}%
\bibitem [{\citenamefont {Duan}\ \emph {et~al.}(2001)\citenamefont {Duan},
  \citenamefont {Lukin}, \citenamefont {Cirac},\ and\ \citenamefont
  {Zoller}}]{Duan2001}%
  \BibitemOpen
  \bibfield  {author} {\bibinfo {author} {\bibfnamefont {L.~M.}\ \bibnamefont
  {Duan}}, \bibinfo {author} {\bibfnamefont {M.~D.}\ \bibnamefont {Lukin}},
  \bibinfo {author} {\bibfnamefont {J.~I.}\ \bibnamefont {Cirac}}, \ and\
  \bibinfo {author} {\bibfnamefont {P.}~\bibnamefont {Zoller}},\ }\href
  {http://dx.doi.org/10.1038/35106500} {\bibfield  {journal} {\bibinfo
  {journal} {Nature}\ }\textbf {\bibinfo {volume} {414}},\ \bibinfo {pages}
  {413} (\bibinfo {year} {2001})}\BibitemShut {NoStop}%
\bibitem [{\citenamefont {Sangouard}\ \emph {et~al.}(2011)\citenamefont
  {Sangouard}, \citenamefont {Simon}, \citenamefont {de~Riedmatten},\ and\
  \citenamefont {Gisin}}]{Sangouard11}%
  \BibitemOpen
  \bibfield  {author} {\bibinfo {author} {\bibfnamefont {N.}~\bibnamefont
  {Sangouard}}, \bibinfo {author} {\bibfnamefont {C.}~\bibnamefont {Simon}},
  \bibinfo {author} {\bibfnamefont {H.}~\bibnamefont {de~Riedmatten}}, \ and\
  \bibinfo {author} {\bibfnamefont {N.}~\bibnamefont {Gisin}},\ }\href
  {\doibase 10.1103/RevModPhys.83.33} {\bibfield  {journal} {\bibinfo
  {journal} {Rev. Mod. Phys.}\ }\textbf {\bibinfo {volume} {83}},\ \bibinfo
  {pages} {33} (\bibinfo {year} {2011})}\BibitemShut {NoStop}%
\bibitem [{\citenamefont {Sangouard}\ \emph {et~al.}(2009)\citenamefont
  {Sangouard}, \citenamefont {Dubessy},\ and\ \citenamefont
  {Simon}}]{Sangouard09}%
  \BibitemOpen
  \bibfield  {author} {\bibinfo {author} {\bibfnamefont {N.}~\bibnamefont
  {Sangouard}}, \bibinfo {author} {\bibfnamefont {R.}~\bibnamefont {Dubessy}},
  \ and\ \bibinfo {author} {\bibfnamefont {C.}~\bibnamefont {Simon}},\ }\href
  {\doibase 10.1103/PhysRevA.79.042340} {\bibfield  {journal} {\bibinfo
  {journal} {Phys. Rev. A}\ }\textbf {\bibinfo {volume} {79}},\ \bibinfo
  {pages} {042340} (\bibinfo {year} {2009})}\BibitemShut {NoStop}%
\bibitem [{\citenamefont {Duan}\ and\ \citenamefont {Monroe}(2010)}]{DuanRMP}%
  \BibitemOpen
  \bibfield  {author} {\bibinfo {author} {\bibfnamefont {L.-M.}\ \bibnamefont
  {Duan}}\ and\ \bibinfo {author} {\bibfnamefont {C.}~\bibnamefont {Monroe}},\
  }\href {\doibase 10.1103/RevModPhys.82.1209} {\bibfield  {journal} {\bibinfo
  {journal} {Rev. Mod. Phys.}\ }\textbf {\bibinfo {volume} {82}},\ \bibinfo
  {pages} {1209} (\bibinfo {year} {2010})}\BibitemShut {NoStop}%
\bibitem [{\citenamefont {Pfister}\ \emph {et~al.}(2016)\citenamefont
  {Pfister}, \citenamefont {Salz}, \citenamefont {Hettrich}, \citenamefont
  {Poschinger},\ and\ \citenamefont {Schmidt-Kaler}}]{Pfister2016}%
  \BibitemOpen
  \bibfield  {author} {\bibinfo {author} {\bibfnamefont {A.~D.}\ \bibnamefont
  {Pfister}}, \bibinfo {author} {\bibfnamefont {M.}~\bibnamefont {Salz}},
  \bibinfo {author} {\bibfnamefont {M.}~\bibnamefont {Hettrich}}, \bibinfo
  {author} {\bibfnamefont {U.~G.}\ \bibnamefont {Poschinger}}, \ and\ \bibinfo
  {author} {\bibfnamefont {F.}~\bibnamefont {Schmidt-Kaler}},\ }\href {\doibase
  10.1007/s00340-016-6362-7} {\bibfield  {journal} {\bibinfo  {journal}
  {Applied Physics B}\ }\textbf {\bibinfo {volume} {122}},\ \bibinfo {pages}
  {89} (\bibinfo {year} {2016})}\BibitemShut {NoStop}%
\bibitem [{\citenamefont {Brunner}\ \emph {et~al.}(2014)\citenamefont
  {Brunner}, \citenamefont {Cavalcanti}, \citenamefont {Pironio}, \citenamefont
  {Scarani},\ and\ \citenamefont {Wehner}}]{BellRMP}%
  \BibitemOpen
  \bibfield  {author} {\bibinfo {author} {\bibfnamefont {N.}~\bibnamefont
  {Brunner}}, \bibinfo {author} {\bibfnamefont {D.}~\bibnamefont {Cavalcanti}},
  \bibinfo {author} {\bibfnamefont {S.}~\bibnamefont {Pironio}}, \bibinfo
  {author} {\bibfnamefont {V.}~\bibnamefont {Scarani}}, \ and\ \bibinfo
  {author} {\bibfnamefont {S.}~\bibnamefont {Wehner}},\ }\href {\doibase
  10.1103/RevModPhys.86.419} {\bibfield  {journal} {\bibinfo  {journal} {Rev.
  Mod. Phys.}\ }\textbf {\bibinfo {volume} {86}},\ \bibinfo {pages} {419}
  (\bibinfo {year} {2014})}\BibitemShut {NoStop}%
\bibitem [{\citenamefont {Zanardi}\ and\ \citenamefont
  {Rasetti}(1997)}]{Zanardi1997}%
  \BibitemOpen
  \bibfield  {author} {\bibinfo {author} {\bibfnamefont {P.}~\bibnamefont
  {Zanardi}}\ and\ \bibinfo {author} {\bibfnamefont {M.}~\bibnamefont
  {Rasetti}},\ }\href {\doibase 10.1103/PhysRevLett.79.3306} {\bibfield
  {journal} {\bibinfo  {journal} {Phys. Rev. Lett.}\ }\textbf {\bibinfo
  {volume} {79}},\ \bibinfo {pages} {3306} (\bibinfo {year}
  {1997})}\BibitemShut {NoStop}%
\bibitem [{\citenamefont {Lidar}\ \emph {et~al.}(1998)\citenamefont {Lidar},
  \citenamefont {Chuang},\ and\ \citenamefont {Whaley}}]{Lidar1998}%
  \BibitemOpen
  \bibfield  {author} {\bibinfo {author} {\bibfnamefont {D.~A.}\ \bibnamefont
  {Lidar}}, \bibinfo {author} {\bibfnamefont {I.~L.}\ \bibnamefont {Chuang}}, \
  and\ \bibinfo {author} {\bibfnamefont {K.~B.}\ \bibnamefont {Whaley}},\
  }\href {\doibase 10.1103/PhysRevLett.81.2594} {\bibfield  {journal} {\bibinfo
   {journal} {Phys. Rev. Lett.}\ }\textbf {\bibinfo {volume} {81}},\ \bibinfo
  {pages} {2594} (\bibinfo {year} {1998})}\BibitemShut {NoStop}%
\bibitem [{\citenamefont {Lita}\ \emph {et~al.}(2008)\citenamefont {Lita},
  \citenamefont {Miller},\ and\ \citenamefont {Nam}}]{detector1}%
  \BibitemOpen
  \bibfield  {author} {\bibinfo {author} {\bibfnamefont {A.~E.}\ \bibnamefont
  {Lita}}, \bibinfo {author} {\bibfnamefont {A.~J.}\ \bibnamefont {Miller}}, \
  and\ \bibinfo {author} {\bibfnamefont {S.~W.}\ \bibnamefont {Nam}},\ }\href
  {\doibase 10.1364/OE.16.003032} {\bibfield  {journal} {\bibinfo  {journal}
  {Opt. Express}\ }\textbf {\bibinfo {volume} {16}},\ \bibinfo {pages} {3032}
  (\bibinfo {year} {2008})}\BibitemShut {NoStop}%
\bibitem [{\citenamefont {Marsili}\ \emph {et~al.}(2013)\citenamefont
  {Marsili}, \citenamefont {Verma}, \citenamefont {Stern}, \citenamefont
  {Harrington}, \citenamefont {Lita}, \citenamefont {Gerrits}, \citenamefont
  {Vayshenker}, \citenamefont {Baek}, \citenamefont {Shaw}, \citenamefont
  {Mirin},\ and\ \citenamefont {Nam}}]{detector2}%
  \BibitemOpen
  \bibfield  {author} {\bibinfo {author} {\bibfnamefont {F.}~\bibnamefont
  {Marsili}}, \bibinfo {author} {\bibfnamefont {V.~B.}\ \bibnamefont {Verma}},
  \bibinfo {author} {\bibfnamefont {J.~A.}\ \bibnamefont {Stern}}, \bibinfo
  {author} {\bibfnamefont {S.}~\bibnamefont {Harrington}}, \bibinfo {author}
  {\bibfnamefont {A.~E.}\ \bibnamefont {Lita}}, \bibinfo {author}
  {\bibfnamefont {T.}~\bibnamefont {Gerrits}}, \bibinfo {author} {\bibfnamefont
  {I.}~\bibnamefont {Vayshenker}}, \bibinfo {author} {\bibfnamefont
  {B.}~\bibnamefont {Baek}}, \bibinfo {author} {\bibfnamefont {M.~D.}\
  \bibnamefont {Shaw}}, \bibinfo {author} {\bibfnamefont {R.~P.}\ \bibnamefont
  {Mirin}}, \ and\ \bibinfo {author} {\bibfnamefont {S.~W.}\ \bibnamefont
  {Nam}},\ }\href {http://dx.doi.org/10.1038/nphoton.2013.13} {\bibfield
  {journal} {\bibinfo  {journal} {Nat Photon}\ }\textbf {\bibinfo {volume}
  {7}},\ \bibinfo {pages} {210} (\bibinfo {year} {2013})}\BibitemShut {NoStop}%
\bibitem [{\citenamefont {Monz}\ \emph {et~al.}(2011)\citenamefont {Monz},
  \citenamefont {Schindler}, \citenamefont {Barreiro}, \citenamefont {Chwalla},
  \citenamefont {Nigg}, \citenamefont {Coish}, \citenamefont {Harlander},
  \citenamefont {H\"ansel}, \citenamefont {Hennrich},\ and\ \citenamefont
  {Blatt}}]{Monz2011}%
  \BibitemOpen
  \bibfield  {author} {\bibinfo {author} {\bibfnamefont {T.}~\bibnamefont
  {Monz}}, \bibinfo {author} {\bibfnamefont {P.}~\bibnamefont {Schindler}},
  \bibinfo {author} {\bibfnamefont {J.~T.}\ \bibnamefont {Barreiro}}, \bibinfo
  {author} {\bibfnamefont {M.}~\bibnamefont {Chwalla}}, \bibinfo {author}
  {\bibfnamefont {D.}~\bibnamefont {Nigg}}, \bibinfo {author} {\bibfnamefont
  {W.~A.}\ \bibnamefont {Coish}}, \bibinfo {author} {\bibfnamefont
  {M.}~\bibnamefont {Harlander}}, \bibinfo {author} {\bibfnamefont
  {W.}~\bibnamefont {H\"ansel}}, \bibinfo {author} {\bibfnamefont
  {M.}~\bibnamefont {Hennrich}}, \ and\ \bibinfo {author} {\bibfnamefont
  {R.}~\bibnamefont {Blatt}},\ }\href {\doibase 10.1103/PhysRevLett.106.130506}
  {\bibfield  {journal} {\bibinfo  {journal} {Phys. Rev. Lett.}\ }\textbf
  {\bibinfo {volume} {106}},\ \bibinfo {pages} {130506} (\bibinfo {year}
  {2011})}\BibitemShut {NoStop}%
\bibitem [{\citenamefont {Van~Meter}(2014)}]{vanmeter2014}%
  \BibitemOpen
  \bibfield  {author} {\bibinfo {author} {\bibfnamefont {R.}~\bibnamefont
  {Van~Meter}},\ }\href {https://books.google.at/books?id=khmNAwAAQBAJ} {\emph
  {\bibinfo {title} {Quantum Networking}}},\ ISTE\ (\bibinfo  {publisher}
  {Wiley},\ \bibinfo {year} {2014})\BibitemShut {NoStop}%
\bibitem [{\citenamefont {Jones}\ \emph {et~al.}(2016)\citenamefont {Jones},
  \citenamefont {Kim}, \citenamefont {Rakher}, \citenamefont {Kwiat},\ and\
  \citenamefont {Ladd}}]{Jones2016}%
  \BibitemOpen
  \bibfield  {author} {\bibinfo {author} {\bibfnamefont {C.}~\bibnamefont
  {Jones}}, \bibinfo {author} {\bibfnamefont {D.}~\bibnamefont {Kim}}, \bibinfo
  {author} {\bibfnamefont {M.~T.}\ \bibnamefont {Rakher}}, \bibinfo {author}
  {\bibfnamefont {P.~G.}\ \bibnamefont {Kwiat}}, \ and\ \bibinfo {author}
  {\bibfnamefont {T.~D.}\ \bibnamefont {Ladd}},\ }\href
  {http://stacks.iop.org/1367-2630/18/i=8/a=083015} {\bibfield  {journal}
  {\bibinfo  {journal} {New Journal of Physics}\ }\textbf {\bibinfo {volume}
  {18}},\ \bibinfo {pages} {083015} (\bibinfo {year} {2016})}\BibitemShut
  {NoStop}%
\bibitem [{\citenamefont {Northup}\ and\ \citenamefont
  {Blatt}(2014)}]{Northup:2014aa}%
  \BibitemOpen
  \bibfield  {author} {\bibinfo {author} {\bibfnamefont {T.~E.}\ \bibnamefont
  {Northup}}\ and\ \bibinfo {author} {\bibfnamefont {R.}~\bibnamefont
  {Blatt}},\ }\href {http://dx.doi.org/10.1038/nphoton.2014.53} {\bibfield
  {journal} {\bibinfo  {journal} {Nat Photon}\ }\textbf {\bibinfo {volume}
  {8}},\ \bibinfo {pages} {356} (\bibinfo {year} {2014})}\BibitemShut {NoStop}%
\bibitem [{\citenamefont {Stute}\ \emph
  {et~al.}(2012{\natexlab{a}})\citenamefont {Stute}, \citenamefont {Casabone},
  \citenamefont {Brandst{\"a}tter}, \citenamefont {Habicher}, \citenamefont
  {Barros}, \citenamefont {Schmidt}, \citenamefont {Northup},\ and\
  \citenamefont {Blatt}}]{Stute2012Appl}%
  \BibitemOpen
  \bibfield  {author} {\bibinfo {author} {\bibfnamefont {A.}~\bibnamefont
  {Stute}}, \bibinfo {author} {\bibfnamefont {B.}~\bibnamefont {Casabone}},
  \bibinfo {author} {\bibfnamefont {B.}~\bibnamefont {Brandst{\"a}tter}},
  \bibinfo {author} {\bibfnamefont {D.}~\bibnamefont {Habicher}}, \bibinfo
  {author} {\bibfnamefont {H.~G.}\ \bibnamefont {Barros}}, \bibinfo {author}
  {\bibfnamefont {P.~O.}\ \bibnamefont {Schmidt}}, \bibinfo {author}
  {\bibfnamefont {T.~E.}\ \bibnamefont {Northup}}, \ and\ \bibinfo {author}
  {\bibfnamefont {R.}~\bibnamefont {Blatt}},\ }\href {\doibase
  10.1007/s00340-011-4861-0} {\bibfield  {journal} {\bibinfo  {journal}
  {Applied Physics B}\ }\textbf {\bibinfo {volume} {107}},\ \bibinfo {pages}
  {1145} (\bibinfo {year} {2012}{\natexlab{a}})}\BibitemShut {NoStop}%
\bibitem [{\citenamefont {Stute}\ \emph
  {et~al.}(2012{\natexlab{b}})\citenamefont {Stute}, \citenamefont {Casabone},
  \citenamefont {Schindler}, \citenamefont {Monz}, \citenamefont {Schmidt},
  \citenamefont {Brandstatter}, \citenamefont {Northup},\ and\ \citenamefont
  {Blatt}}]{Stute2012}%
  \BibitemOpen
  \bibfield  {author} {\bibinfo {author} {\bibfnamefont {A.}~\bibnamefont
  {Stute}}, \bibinfo {author} {\bibfnamefont {B.}~\bibnamefont {Casabone}},
  \bibinfo {author} {\bibfnamefont {P.}~\bibnamefont {Schindler}}, \bibinfo
  {author} {\bibfnamefont {T.}~\bibnamefont {Monz}}, \bibinfo {author}
  {\bibfnamefont {P.~O.}\ \bibnamefont {Schmidt}}, \bibinfo {author}
  {\bibfnamefont {B.}~\bibnamefont {Brandstatter}}, \bibinfo {author}
  {\bibfnamefont {T.~E.}\ \bibnamefont {Northup}}, \ and\ \bibinfo {author}
  {\bibfnamefont {R.}~\bibnamefont {Blatt}},\ }\href
  {http://dx.doi.org/10.1038/nature11120} {\bibfield  {journal} {\bibinfo
  {journal} {Nature}\ }\textbf {\bibinfo {volume} {485}},\ \bibinfo {pages}
  {482} (\bibinfo {year} {2012}{\natexlab{b}})}\BibitemShut {NoStop}%
\bibitem [{\citenamefont {Schindler}\ \emph {et~al.}(2013)\citenamefont
  {Schindler}, \citenamefont {Nigg}, \citenamefont {Monz}, \citenamefont
  {Barreiro}, \citenamefont {Martinez}, \citenamefont {Wang}, \citenamefont
  {Quint}, \citenamefont {Brandl}, \citenamefont {Nebendahl}, \citenamefont
  {Roos}, \citenamefont {Chwalla}, \citenamefont {Hennrich},\ and\
  \citenamefont {Blatt}}]{Schindler2013}%
  \BibitemOpen
  \bibfield  {author} {\bibinfo {author} {\bibfnamefont {P.}~\bibnamefont
  {Schindler}}, \bibinfo {author} {\bibfnamefont {D.}~\bibnamefont {Nigg}},
  \bibinfo {author} {\bibfnamefont {T.}~\bibnamefont {Monz}}, \bibinfo {author}
  {\bibfnamefont {J.~T.}\ \bibnamefont {Barreiro}}, \bibinfo {author}
  {\bibfnamefont {E.}~\bibnamefont {Martinez}}, \bibinfo {author}
  {\bibfnamefont {S.~X.}\ \bibnamefont {Wang}}, \bibinfo {author}
  {\bibfnamefont {S.}~\bibnamefont {Quint}}, \bibinfo {author} {\bibfnamefont
  {M.~F.}\ \bibnamefont {Brandl}}, \bibinfo {author} {\bibfnamefont
  {V.}~\bibnamefont {Nebendahl}}, \bibinfo {author} {\bibfnamefont {C.~F.}\
  \bibnamefont {Roos}}, \bibinfo {author} {\bibfnamefont {M.}~\bibnamefont
  {Chwalla}}, \bibinfo {author} {\bibfnamefont {M.}~\bibnamefont {Hennrich}}, \
  and\ \bibinfo {author} {\bibfnamefont {R.}~\bibnamefont {Blatt}},\ }\href
  {http://stacks.iop.org/1367-2630/15/i=12/a=123012} {\bibfield  {journal}
  {\bibinfo  {journal} {New Journal of Physics}\ }\textbf {\bibinfo {volume}
  {15}},\ \bibinfo {pages} {123012} (\bibinfo {year} {2013})}\BibitemShut
  {NoStop}%
\bibitem [{\citenamefont {Bennett}\ \emph
  {et~al.}(1996{\natexlab{a}})\citenamefont {Bennett}, \citenamefont
  {Brassard}, \citenamefont {Popescu}, \citenamefont {Schumacher},
  \citenamefont {Smolin},\ and\ \citenamefont {Wootters}}]{Bennett96}%
  \BibitemOpen
  \bibfield  {author} {\bibinfo {author} {\bibfnamefont {C.~H.}\ \bibnamefont
  {Bennett}}, \bibinfo {author} {\bibfnamefont {G.}~\bibnamefont {Brassard}},
  \bibinfo {author} {\bibfnamefont {S.}~\bibnamefont {Popescu}}, \bibinfo
  {author} {\bibfnamefont {B.}~\bibnamefont {Schumacher}}, \bibinfo {author}
  {\bibfnamefont {J.~A.}\ \bibnamefont {Smolin}}, \ and\ \bibinfo {author}
  {\bibfnamefont {W.~K.}\ \bibnamefont {Wootters}},\ }\href {\doibase
  10.1103/PhysRevLett.76.722} {\bibfield  {journal} {\bibinfo  {journal} {Phys.
  Rev. Lett.}\ }\textbf {\bibinfo {volume} {76}},\ \bibinfo {pages} {722}
  (\bibinfo {year} {1996}{\natexlab{a}})}\BibitemShut {NoStop}%
\bibitem [{\citenamefont {Zaske}\ \emph {et~al.}(2012)\citenamefont {Zaske},
  \citenamefont {Lenhard}, \citenamefont {Ke\ss{}ler}, \citenamefont {Kettler},
  \citenamefont {Hepp}, \citenamefont {Arend}, \citenamefont {Albrecht},
  \citenamefont {Schulz}, \citenamefont {Jetter}, \citenamefont {Michler},\
  and\ \citenamefont {Becher}}]{Zaske2012}%
  \BibitemOpen
  \bibfield  {author} {\bibinfo {author} {\bibfnamefont {S.}~\bibnamefont
  {Zaske}}, \bibinfo {author} {\bibfnamefont {A.}~\bibnamefont {Lenhard}},
  \bibinfo {author} {\bibfnamefont {C.~A.}\ \bibnamefont {Ke\ss{}ler}},
  \bibinfo {author} {\bibfnamefont {J.}~\bibnamefont {Kettler}}, \bibinfo
  {author} {\bibfnamefont {C.}~\bibnamefont {Hepp}}, \bibinfo {author}
  {\bibfnamefont {C.}~\bibnamefont {Arend}}, \bibinfo {author} {\bibfnamefont
  {R.}~\bibnamefont {Albrecht}}, \bibinfo {author} {\bibfnamefont {W.-M.}\
  \bibnamefont {Schulz}}, \bibinfo {author} {\bibfnamefont {M.}~\bibnamefont
  {Jetter}}, \bibinfo {author} {\bibfnamefont {P.}~\bibnamefont {Michler}}, \
  and\ \bibinfo {author} {\bibfnamefont {C.}~\bibnamefont {Becher}},\ }\href
  {\doibase 10.1103/PhysRevLett.109.147404} {\bibfield  {journal} {\bibinfo
  {journal} {Phys. Rev. Lett.}\ }\textbf {\bibinfo {volume} {109}},\ \bibinfo
  {pages} {147404} (\bibinfo {year} {2012})}\BibitemShut {NoStop}%
\bibitem [{\citenamefont {Albrecht}\ \emph {et~al.}(2014)\citenamefont
  {Albrecht}, \citenamefont {Farrera}, \citenamefont {Fernandez-Gonzalvo},
  \citenamefont {Cristiani},\ and\ \citenamefont
  {de~Riedmatten}}]{Albrecht2014}%
  \BibitemOpen
  \bibfield  {author} {\bibinfo {author} {\bibfnamefont {B.}~\bibnamefont
  {Albrecht}}, \bibinfo {author} {\bibfnamefont {P.}~\bibnamefont {Farrera}},
  \bibinfo {author} {\bibfnamefont {X.}~\bibnamefont {Fernandez-Gonzalvo}},
  \bibinfo {author} {\bibfnamefont {M.}~\bibnamefont {Cristiani}}, \ and\
  \bibinfo {author} {\bibfnamefont {H.}~\bibnamefont {de~Riedmatten}},\ }\href
  {http://dx.doi.org/10.1038/ncomms4376} {\bibfield  {journal} {\bibinfo
  {journal} {Nature Communications}\ }\textbf {\bibinfo {volume} {5}},\
  \bibinfo {pages} {3376 EP } (\bibinfo {year} {2014})}\BibitemShut {NoStop}%
\bibitem [{\citenamefont {Pelc}\ \emph {et~al.}(2011)\citenamefont {Pelc},
  \citenamefont {Ma}, \citenamefont {Phillips}, \citenamefont {Zhang},
  \citenamefont {Langrock}, \citenamefont {Slattery}, \citenamefont {Tang},\
  and\ \citenamefont {Fejer}}]{Pelc2011}%
  \BibitemOpen
  \bibfield  {author} {\bibinfo {author} {\bibfnamefont {J.~S.}\ \bibnamefont
  {Pelc}}, \bibinfo {author} {\bibfnamefont {L.}~\bibnamefont {Ma}}, \bibinfo
  {author} {\bibfnamefont {C.~R.}\ \bibnamefont {Phillips}}, \bibinfo {author}
  {\bibfnamefont {Q.}~\bibnamefont {Zhang}}, \bibinfo {author} {\bibfnamefont
  {C.}~\bibnamefont {Langrock}}, \bibinfo {author} {\bibfnamefont
  {O.}~\bibnamefont {Slattery}}, \bibinfo {author} {\bibfnamefont
  {X.}~\bibnamefont {Tang}}, \ and\ \bibinfo {author} {\bibfnamefont {M.~M.}\
  \bibnamefont {Fejer}},\ }\href {\doibase 10.1364/OE.19.021445} {\bibfield
  {journal} {\bibinfo  {journal} {Opt. Express}\ }\textbf {\bibinfo {volume}
  {19}},\ \bibinfo {pages} {21445} (\bibinfo {year} {2011})}\BibitemShut
  {NoStop}%
\bibitem [{\citenamefont {Casabone}\ \emph {et~al.}(2013)\citenamefont
  {Casabone}, \citenamefont {Stute}, \citenamefont {Friebe}, \citenamefont
  {Brandst\"atter}, \citenamefont {Sch\"uppert}, \citenamefont {Blatt},\ and\
  \citenamefont {Northup}}]{Casabone2013}%
  \BibitemOpen
  \bibfield  {author} {\bibinfo {author} {\bibfnamefont {B.}~\bibnamefont
  {Casabone}}, \bibinfo {author} {\bibfnamefont {A.}~\bibnamefont {Stute}},
  \bibinfo {author} {\bibfnamefont {K.}~\bibnamefont {Friebe}}, \bibinfo
  {author} {\bibfnamefont {B.}~\bibnamefont {Brandst\"atter}}, \bibinfo
  {author} {\bibfnamefont {K.}~\bibnamefont {Sch\"uppert}}, \bibinfo {author}
  {\bibfnamefont {R.}~\bibnamefont {Blatt}}, \ and\ \bibinfo {author}
  {\bibfnamefont {T.~E.}\ \bibnamefont {Northup}},\ }\href {\doibase
  10.1103/PhysRevLett.111.100505} {\bibfield  {journal} {\bibinfo  {journal}
  {Phys. Rev. Lett.}\ }\textbf {\bibinfo {volume} {111}},\ \bibinfo {pages}
  {100505} (\bibinfo {year} {2013})}\BibitemShut {NoStop}%
\bibitem [{\citenamefont {Lechner}\ \emph {et~al.}(2016)\citenamefont
  {Lechner}, \citenamefont {Maier}, \citenamefont {Hempel}, \citenamefont
  {Jurcevic}, \citenamefont {Lanyon}, \citenamefont {Monz}, \citenamefont
  {Brownnutt}, \citenamefont {Blatt},\ and\ \citenamefont
  {Roos}}]{Lechner2016}%
  \BibitemOpen
  \bibfield  {author} {\bibinfo {author} {\bibfnamefont {R.}~\bibnamefont
  {Lechner}}, \bibinfo {author} {\bibfnamefont {C.}~\bibnamefont {Maier}},
  \bibinfo {author} {\bibfnamefont {C.}~\bibnamefont {Hempel}}, \bibinfo
  {author} {\bibfnamefont {P.}~\bibnamefont {Jurcevic}}, \bibinfo {author}
  {\bibfnamefont {B.~P.}\ \bibnamefont {Lanyon}}, \bibinfo {author}
  {\bibfnamefont {T.}~\bibnamefont {Monz}}, \bibinfo {author} {\bibfnamefont
  {M.}~\bibnamefont {Brownnutt}}, \bibinfo {author} {\bibfnamefont
  {R.}~\bibnamefont {Blatt}}, \ and\ \bibinfo {author} {\bibfnamefont {C.~F.}\
  \bibnamefont {Roos}},\ }\href {\doibase 10.1103/PhysRevA.93.053401}
  {\bibfield  {journal} {\bibinfo  {journal} {Phys. Rev. A}\ }\textbf {\bibinfo
  {volume} {93}},\ \bibinfo {pages} {053401} (\bibinfo {year}
  {2016})}\BibitemShut {NoStop}%
\bibitem [{\citenamefont {Roos}\ \emph {et~al.}(2004)\citenamefont {Roos},
  \citenamefont {Lancaster}, \citenamefont {Riebe}, \citenamefont {H\"affner},
  \citenamefont {H\"ansel}, \citenamefont {Gulde}, \citenamefont {Becher},
  \citenamefont {Eschner}, \citenamefont {Schmidt-Kaler},\ and\ \citenamefont
  {Blatt}}]{Roos2004}%
  \BibitemOpen
  \bibfield  {author} {\bibinfo {author} {\bibfnamefont {C.~F.}\ \bibnamefont
  {Roos}}, \bibinfo {author} {\bibfnamefont {G.~P.~T.}\ \bibnamefont
  {Lancaster}}, \bibinfo {author} {\bibfnamefont {M.}~\bibnamefont {Riebe}},
  \bibinfo {author} {\bibfnamefont {H.}~\bibnamefont {H\"affner}}, \bibinfo
  {author} {\bibfnamefont {W.}~\bibnamefont {H\"ansel}}, \bibinfo {author}
  {\bibfnamefont {S.}~\bibnamefont {Gulde}}, \bibinfo {author} {\bibfnamefont
  {C.}~\bibnamefont {Becher}}, \bibinfo {author} {\bibfnamefont
  {J.}~\bibnamefont {Eschner}}, \bibinfo {author} {\bibfnamefont
  {F.}~\bibnamefont {Schmidt-Kaler}}, \ and\ \bibinfo {author} {\bibfnamefont
  {R.}~\bibnamefont {Blatt}},\ }\href {\doibase 10.1103/PhysRevLett.92.220402}
  {\bibfield  {journal} {\bibinfo  {journal} {Phys. Rev. Lett.}\ }\textbf
  {\bibinfo {volume} {92}},\ \bibinfo {pages} {220402} (\bibinfo {year}
  {2004})}\BibitemShut {NoStop}%
\bibitem [{\citenamefont {H{\"a}ffner}\ \emph {et~al.}(2005)\citenamefont
  {H{\"a}ffner}, \citenamefont {Schmidt-Kaler}, \citenamefont {H{\"a}nsel},
  \citenamefont {Roos}, \citenamefont {K{\"o}rber}, \citenamefont {Chwalla},
  \citenamefont {Riebe}, \citenamefont {Benhelm}, \citenamefont {Rapol},
  \citenamefont {Becher},\ and\ \citenamefont {Blatt}}]{Haffner2005}%
  \BibitemOpen
  \bibfield  {author} {\bibinfo {author} {\bibfnamefont {H.}~\bibnamefont
  {H{\"a}ffner}}, \bibinfo {author} {\bibfnamefont {F.}~\bibnamefont
  {Schmidt-Kaler}}, \bibinfo {author} {\bibfnamefont {W.}~\bibnamefont
  {H{\"a}nsel}}, \bibinfo {author} {\bibfnamefont {C.~F.}\ \bibnamefont
  {Roos}}, \bibinfo {author} {\bibfnamefont {T.}~\bibnamefont {K{\"o}rber}},
  \bibinfo {author} {\bibfnamefont {M.}~\bibnamefont {Chwalla}}, \bibinfo
  {author} {\bibfnamefont {M.}~\bibnamefont {Riebe}}, \bibinfo {author}
  {\bibfnamefont {J.}~\bibnamefont {Benhelm}}, \bibinfo {author} {\bibfnamefont
  {U.~D.}\ \bibnamefont {Rapol}}, \bibinfo {author} {\bibfnamefont
  {C.}~\bibnamefont {Becher}}, \ and\ \bibinfo {author} {\bibfnamefont
  {R.}~\bibnamefont {Blatt}},\ }\href {\doibase 10.1007/s00340-005-1917-z}
  {\bibfield  {journal} {\bibinfo  {journal} {Applied Physics B}\ }\textbf
  {\bibinfo {volume} {81}},\ \bibinfo {pages} {151} (\bibinfo {year}
  {2005})}\BibitemShut {NoStop}%
\bibitem [{\citenamefont {Bell}(1964)}]{Bell}%
  \BibitemOpen
  \bibfield  {author} {\bibinfo {author} {\bibfnamefont {J.~S.}\ \bibnamefont
  {Bell}},\ }\href@noop {} {\bibfield  {journal} {\bibinfo  {journal}
  {Physics}\ }\textbf {\bibinfo {volume} {1}},\ \bibinfo {pages} {195 }
  (\bibinfo {year} {1964})}\BibitemShut {NoStop}%
\bibitem [{\citenamefont {Clauser}\ \emph {et~al.}(1969)\citenamefont
  {Clauser}, \citenamefont {Horne}, \citenamefont {Shimony},\ and\
  \citenamefont {Holt}}]{CHSH}%
  \BibitemOpen
  \bibfield  {author} {\bibinfo {author} {\bibfnamefont {J.~F.}\ \bibnamefont
  {Clauser}}, \bibinfo {author} {\bibfnamefont {M.~A.}\ \bibnamefont {Horne}},
  \bibinfo {author} {\bibfnamefont {A.}~\bibnamefont {Shimony}}, \ and\
  \bibinfo {author} {\bibfnamefont {R.~A.}\ \bibnamefont {Holt}},\ }\href
  {\doibase 10.1103/PhysRevLett.23.880} {\bibfield  {journal} {\bibinfo
  {journal} {Phys. Rev. Lett.}\ }\textbf {\bibinfo {volume} {23}},\ \bibinfo
  {pages} {880} (\bibinfo {year} {1969})}\BibitemShut {NoStop}%
\bibitem [{\citenamefont {Bennett}\ \emph
  {et~al.}(1996{\natexlab{b}})\citenamefont {Bennett}, \citenamefont
  {Brassard}, \citenamefont {Popescu}, \citenamefont {Schumacher},
  \citenamefont {Smolin},\ and\ \citenamefont {Wootters}}]{Be96}%
  \BibitemOpen
  \bibfield  {author} {\bibinfo {author} {\bibfnamefont {C.~H.}\ \bibnamefont
  {Bennett}}, \bibinfo {author} {\bibfnamefont {G.}~\bibnamefont {Brassard}},
  \bibinfo {author} {\bibfnamefont {S.}~\bibnamefont {Popescu}}, \bibinfo
  {author} {\bibfnamefont {B.}~\bibnamefont {Schumacher}}, \bibinfo {author}
  {\bibfnamefont {J.~A.}\ \bibnamefont {Smolin}}, \ and\ \bibinfo {author}
  {\bibfnamefont {W.~K.}\ \bibnamefont {Wootters}},\ }\href {\doibase
  10.1103/PhysRevLett.76.722} {\bibfield  {journal} {\bibinfo  {journal} {Phys.
  Rev. Lett.}\ }\textbf {\bibinfo {volume} {76}},\ \bibinfo {pages} {722}
  (\bibinfo {year} {1996}{\natexlab{b}})}\BibitemShut {NoStop}%
\bibitem [{\citenamefont {Collins}\ \emph {et~al.}(2007)\citenamefont
  {Collins}, \citenamefont {Jenkins}, \citenamefont {Kuzmich},\ and\
  \citenamefont {Kennedy}}]{Collins07}%
  \BibitemOpen
  \bibfield  {author} {\bibinfo {author} {\bibfnamefont {O.~A.}\ \bibnamefont
  {Collins}}, \bibinfo {author} {\bibfnamefont {S.~D.}\ \bibnamefont
  {Jenkins}}, \bibinfo {author} {\bibfnamefont {A.}~\bibnamefont {Kuzmich}}, \
  and\ \bibinfo {author} {\bibfnamefont {T.~A.~B.}\ \bibnamefont {Kennedy}},\
  }\href {\doibase 10.1103/PhysRevLett.98.060502} {\bibfield  {journal}
  {\bibinfo  {journal} {Phys. Rev. Lett.}\ }\textbf {\bibinfo {volume} {98}},\
  \bibinfo {pages} {060502} (\bibinfo {year} {2007})}\BibitemShut {NoStop}%
\bibitem [{\citenamefont {Ballance}\ \emph {et~al.}(2016)\citenamefont
  {Ballance}, \citenamefont {Harty}, \citenamefont {Linke}, \citenamefont
  {Sepiol},\ and\ \citenamefont {Lucas}}]{Ballance15}%
  \BibitemOpen
  \bibfield  {author} {\bibinfo {author} {\bibfnamefont {C.~J.}\ \bibnamefont
  {Ballance}}, \bibinfo {author} {\bibfnamefont {T.~P.}\ \bibnamefont {Harty}},
  \bibinfo {author} {\bibfnamefont {N.~M.}\ \bibnamefont {Linke}}, \bibinfo
  {author} {\bibfnamefont {M.~A.}\ \bibnamefont {Sepiol}}, \ and\ \bibinfo
  {author} {\bibfnamefont {D.~M.}\ \bibnamefont {Lucas}},\ }\href {\doibase
  10.1103/PhysRevLett.117.060504} {\bibfield  {journal} {\bibinfo  {journal}
  {Phys. Rev. Lett.}\ }\textbf {\bibinfo {volume} {117}},\ \bibinfo {pages}
  {060504} (\bibinfo {year} {2016})}\BibitemShut {NoStop}%
\bibitem [{\citenamefont {Bennett}\ \emph
  {et~al.}(1996{\natexlab{c}})\citenamefont {Bennett}, \citenamefont
  {DiVincenzo}, \citenamefont {Smolin},\ and\ \citenamefont
  {Wootters}}]{Bennett96a}%
  \BibitemOpen
  \bibfield  {author} {\bibinfo {author} {\bibfnamefont {C.~H.}\ \bibnamefont
  {Bennett}}, \bibinfo {author} {\bibfnamefont {D.~P.}\ \bibnamefont
  {DiVincenzo}}, \bibinfo {author} {\bibfnamefont {J.~A.}\ \bibnamefont
  {Smolin}}, \ and\ \bibinfo {author} {\bibfnamefont {W.~K.}\ \bibnamefont
  {Wootters}},\ }\href {\doibase 10.1103/PhysRevA.54.3824} {\bibfield
  {journal} {\bibinfo  {journal} {Phys. Rev. A}\ }\textbf {\bibinfo {volume}
  {54}},\ \bibinfo {pages} {3824} (\bibinfo {year}
  {1996}{\natexlab{c}})}\BibitemShut {NoStop}%
\bibitem [{\citenamefont {Deutsch}\ \emph {et~al.}(1998)\citenamefont
  {Deutsch}, \citenamefont {Ekert}, \citenamefont {Jozsa}, \citenamefont
  {Macchiavello}, \citenamefont {Popescu},\ and\ \citenamefont
  {Sanpera}}]{Deutsch96}%
  \BibitemOpen
  \bibfield  {author} {\bibinfo {author} {\bibfnamefont {D.}~\bibnamefont
  {Deutsch}}, \bibinfo {author} {\bibfnamefont {A.}~\bibnamefont {Ekert}},
  \bibinfo {author} {\bibfnamefont {R.}~\bibnamefont {Jozsa}}, \bibinfo
  {author} {\bibfnamefont {C.}~\bibnamefont {Macchiavello}}, \bibinfo {author}
  {\bibfnamefont {S.}~\bibnamefont {Popescu}}, \ and\ \bibinfo {author}
  {\bibfnamefont {A.}~\bibnamefont {Sanpera}},\ }\href {\doibase
  10.1103/PhysRevLett.80.2022} {\bibfield  {journal} {\bibinfo  {journal}
  {Phys. Rev. Lett.}\ }\textbf {\bibinfo {volume} {80}},\ \bibinfo {pages}
  {2022} (\bibinfo {year} {1998})}\BibitemShut {NoStop}%
\bibitem [{\citenamefont {Klein}\ \emph {et~al.}(2006)\citenamefont {Klein},
  \citenamefont {Dorner}, \citenamefont {Alves},\ and\ \citenamefont
  {Jaksch}}]{Klein2006}%
  \BibitemOpen
  \bibfield  {author} {\bibinfo {author} {\bibfnamefont {A.}~\bibnamefont
  {Klein}}, \bibinfo {author} {\bibfnamefont {U.}~\bibnamefont {Dorner}},
  \bibinfo {author} {\bibfnamefont {C.~M.}\ \bibnamefont {Alves}}, \ and\
  \bibinfo {author} {\bibfnamefont {D.}~\bibnamefont {Jaksch}},\ }\href
  {\doibase 10.1103/PhysRevA.73.012332} {\bibfield  {journal} {\bibinfo
  {journal} {Phys. Rev. A}\ }\textbf {\bibinfo {volume} {73}},\ \bibinfo
  {pages} {012332} (\bibinfo {year} {2006})}\BibitemShut {NoStop}%
\bibitem [{\citenamefont {Dorner}\ \emph {et~al.}(2008)\citenamefont {Dorner},
  \citenamefont {Klein},\ and\ \citenamefont {Jaksch}}]{Dorner2008}%
  \BibitemOpen
  \bibfield  {author} {\bibinfo {author} {\bibfnamefont {U.}~\bibnamefont
  {Dorner}}, \bibinfo {author} {\bibfnamefont {A.}~\bibnamefont {Klein}}, \
  and\ \bibinfo {author} {\bibfnamefont {D.}~\bibnamefont {Jaksch}},\
  }\href@noop {} {\bibfield  {journal} {\bibinfo  {journal} {Quant. Inf.
  Comp.}\ }\textbf {\bibinfo {volume} {8}},\ \bibinfo {pages} {0468} (\bibinfo
  {year} {2008})}\BibitemShut {NoStop}%
\bibitem [{\citenamefont {Reiserer}\ \emph {et~al.}(2016)\citenamefont
  {Reiserer}, \citenamefont {Kalb}, \citenamefont {Blok}, \citenamefont {van
  Bemmelen}, \citenamefont {Taminiau}, \citenamefont {Hanson}, \citenamefont
  {Twitchen},\ and\ \citenamefont {Markham}}]{Reiserer16}%
  \BibitemOpen
  \bibfield  {author} {\bibinfo {author} {\bibfnamefont {A.}~\bibnamefont
  {Reiserer}}, \bibinfo {author} {\bibfnamefont {N.}~\bibnamefont {Kalb}},
  \bibinfo {author} {\bibfnamefont {M.~S.}\ \bibnamefont {Blok}}, \bibinfo
  {author} {\bibfnamefont {K.~J.~M.}\ \bibnamefont {van Bemmelen}}, \bibinfo
  {author} {\bibfnamefont {T.~H.}\ \bibnamefont {Taminiau}}, \bibinfo {author}
  {\bibfnamefont {R.}~\bibnamefont {Hanson}}, \bibinfo {author} {\bibfnamefont
  {D.~J.}\ \bibnamefont {Twitchen}}, \ and\ \bibinfo {author} {\bibfnamefont
  {M.}~\bibnamefont {Markham}},\ }\href {\doibase 10.1103/PhysRevX.6.021040}
  {\bibfield  {journal} {\bibinfo  {journal} {Phys. Rev. X}\ }\textbf {\bibinfo
  {volume} {6}},\ \bibinfo {pages} {021040} (\bibinfo {year}
  {2016})}\BibitemShut {NoStop}%
\bibitem [{\citenamefont {Zwerger}\ \emph {et~al.}(2014)\citenamefont
  {Zwerger}, \citenamefont {Briegel},\ and\ \citenamefont {D{\"u}r}}]{Zw14}%
  \BibitemOpen
  \bibfield  {author} {\bibinfo {author} {\bibfnamefont {M.}~\bibnamefont
  {Zwerger}}, \bibinfo {author} {\bibfnamefont {H.-J.}\ \bibnamefont
  {Briegel}}, \ and\ \bibinfo {author} {\bibfnamefont {W.}~\bibnamefont
  {D{\"u}r}},\ }\href {http://www.ncbi.nlm.nih.gov/pmc/articles/PMC4064337/}
  {\bibfield  {journal} {\bibinfo  {journal} {Scientific Reports}\ }\textbf
  {\bibinfo {volume} {4}},\ \bibinfo {pages} {5364} (\bibinfo {year}
  {2014})}\BibitemShut {NoStop}%
\bibitem [{\citenamefont {Muralidharan}\ \emph {et~al.}(2014)\citenamefont
  {Muralidharan}, \citenamefont {Kim}, \citenamefont {L\"utkenhaus},
  \citenamefont {Lukin},\ and\ \citenamefont {Jiang}}]{Mur14}%
  \BibitemOpen
  \bibfield  {author} {\bibinfo {author} {\bibfnamefont {S.}~\bibnamefont
  {Muralidharan}}, \bibinfo {author} {\bibfnamefont {J.}~\bibnamefont {Kim}},
  \bibinfo {author} {\bibfnamefont {N.}~\bibnamefont {L\"utkenhaus}}, \bibinfo
  {author} {\bibfnamefont {M.~D.}\ \bibnamefont {Lukin}}, \ and\ \bibinfo
  {author} {\bibfnamefont {L.}~\bibnamefont {Jiang}},\ }\href {\doibase
  10.1103/PhysRevLett.112.250501} {\bibfield  {journal} {\bibinfo  {journal}
  {Phys. Rev. Lett.}\ }\textbf {\bibinfo {volume} {112}},\ \bibinfo {pages}
  {250501} (\bibinfo {year} {2014})}\BibitemShut {NoStop}%
\end{thebibliography}%

\end{document}